\newcommand{\mincir}{\raise
-2.truept\hbox{\rlap{\hbox{$\sim$}}\raise5.truept 
\hbox{$<$}\ }}
\newcommand{\magcir}{\raise
-2.truept\hbox{\rlap{\hbox{$\sim$}}\raise5.truept
\hbox{$>$}\ }}
\newcommand{\minmag}{\raise-2.truept\hbox{\rlap{\hbox{$<$}}\raise
6.truept\hbox
{$>$}\ }}

\newcommand{\be}{\begin{equation}}
\newcommand{\ee}{\end{equation}}
\newcommand{\ba}{\begin{eqnarray}}
\newcommand{\ea}{\end{eqnarray}}
\newcommand{\brr}{\begin{array}}
 
\newcommand{\err}{\end{array}}
\newcommand{\bc}{\begin{center}}
\newcommand{\ec}{\end{center}}

\newcommand{\tu}{\textunderscore}

\documentclass{pasa}%

\usepackage[authoryear]{natbib}
\bibpunct{(}{)}{;}{a}{}{,}
\usepackage{graphicx}
\usepackage{txfonts}
\usepackage{amssymb}
\usepackage{times}
\usepackage{wasysym}
\usepackage{threeparttable}
\usepackage{tablefootnote}
\usepackage{footnote}
\usepackage[bottom]{footmisc}
\usepackage{tablefootnote}

\bibpunct{(}{)}{;}{a}{}{,} 

\title[The {\rm SFR-${\rm M}_{\star}$} relation of $z \sim 1-4$ galaxies]{The relation between star formation rate and stellar mass of galaxies at  $z\sim1-4$}
\author[Katsianis et al.]{A. Katsianis\,$^{1,\, 3}$\thanks{kata@das.uchile.cl}, E. Tescari\,$^{2,\, 3}$, \and J. S. B. Wyithe\,$^{2,\, 3}$\\
\affil{$^1$Department of Astronomy, Universitad de Chile, Camino El Observatorio 1515, Las Condes, Santiago, Chile}%
\affil{$^2$School of Physics, The University of Melbourne, Parkville, VIC 3010, Australia}%
\affil{$^3$ARC Centre of Excellence for All-Sky Astrophysics (CAASTRO)}}%
\jid{PASA}
\doi{10.1017/pas.\the\year.xxx}
\jyear{\the\year}

\begin{document}%
\begin{abstract}
The relation between the Star Formation Rate (SFR) and stellar mass (${\rm M}_{\star}$) of galaxies represents a fundamental constraint on galaxy formation, and has been studied extensively both in observations and cosmological hydrodynamic simulations. However, the observed amplitude of the star formation rate - stellar mass relation has not been successfully reproduced in simulations,  indicating either that the halo accretion history and baryonic physics are poorly understood/modeled or that observations contain biases. In this paper, we examine the evolution of the SFR$-{\rm M}_{\star}$ relation of $z\sim 1-4 $ galaxies and display the inconsistency between observed relations that are obtained using different techniques. We employ cosmological hydrodynamic simulations from various groups which are tuned to reproduce a range of observables and compare these with a range of observed SFR$-{\rm M}_{\star}$ relations. We find that numerical results are consistent with observations that use Spectral Energy Distribution (SED) techniques to estimate star formation rates, dust corrections and stellar masses. On the contrary, simulations are not able to reproduce results that were obtained by combining only UV and IR luminosities (UV+IR). These imply SFRs at a fixed stellar mass that are larger almost by a factor of 5 than those of SED measurements for $z \sim 1.5-4$. For $z < 1.5$,  the results from simulations, SED fitting techniques and IR+UV conversion agree well. We find that surveys that preferably select star forming galaxies (e.g. by adopting Lyman-break or blue selection) typically predict a larger median/average star formation rate at a fixed stellar mass especially for high mass objects, with respect to mass selected samples and hydrodynamic simulations. Furthermore, we find remarkable agreement between the numerical results from various authors who have employed different cosmological codes and run simulations with different resolutions. This is interesting for two reasons. A) simulations can produce realistic populations of galaxies within representative cosmological volumes even at relatively modest resolutions. B) It is likely that current numerical codes that rely on similar subgrid multiphase Inter-Stellar Medium (ISM) models and are tuned to reproduce statistical properties of galaxies, produce similar results for the SFR$-{\rm M}_{\star}$ relation by construction, regardless of resolution, box size and, to some extent, the adopted feedback prescriptions.
\end{abstract}
\maketitle%

\section{Introduction}

Numerous studies have demonstrated a correlation between galaxy star formation rates and stellar masses at redshifts $z \sim1-4$. These studies have employed a range of different observing strategies to sample the galaxy population as completely as possible \citep{Noeske2007,Elbaz07,Daddi2007,Drory08,Kajisawa2010,Karim2011,Bauer11,Gilbank2011,reddy2012,bouwens2012,Heinis2014,Koyama2013,Guo2013,deBarros,Speagle14}. Galaxies in different surveys and redshifts are usually selected using different techniques and  wavelengths. Some examples of selecting galaxies in the literature are:

\begin{enumerate}

\item Lyman-break selection: in the absence of dust extinction, star forming galaxies have a flat continuum at rest frame ultraviolet wavelengths. However, blueward of the Lyman break (rest-frame 912 $\AA$), photoelectric absorption by galactic or extragalactic sources of neutral hydrogen sharply cuts the emitted spectrum. High redshift sources can be selected using this spectral break. This method has been extensively used to construct samples of galaxies at different epochs and study their Star Formation Rate-Stellar Mass (SFR-${\rm M}_{\star}$) relation \citep{reddy2012,bouwens2012,Heinis2014,deBarros}. An important drawback of this technique is that it is only capable of selecting objects with young ages, high star formation rates and low dust content (i.e star forming galaxies). Only these objects are able to produce a large amount of UV light which is not absorbed by dust. This makes the Lyman-break technique insensitive  to galaxies that are dust-reddened or contain evolved stellar populations.  \\

\item H$\alpha$ selection: in this case, galaxies are selected by their emission in the H$\alpha$ line (rest-frame 6563 $\AA$), which is correlated with star formation.  Various authors have constructed H$\alpha$ selected samples to study the evolution of the cosmic star formation rate density and SFR$-{\rm M}_{\star}$ relation \citep[e.g.][]{Sobral2013,Koyama2013,Delosrayes2014}. \\

\item K-band selection: The K-luminosity of galaxies measures mass in old stars and therefore is a robust estimator of galaxy stellar mass \citep{Broad92,Brinchmann00}. In this case, galaxies are selected by their luminosity in this band and then the redshift is estimated using Spectral Energy Distribution (SED) fitting \citep{Drory08}. However, the  K-band magnitude limit of surveys can restrict the stellar mass distribution at the low mass end. This is one of the drawbacks of a magnitude-limited survey \citep{reddy2012}. \citet{Drory2005} showed that galaxies can also be selected by mass using I-band, yielding mass functions that are consistent with K-band selection.

\end{enumerate}
Besides the above selection methods that are typically used to construct parent samples of galaxies,  observers adopt further selection criteria to create a sub-population within the initial sample with the   desired properties and exclude other objects. For example, color cuts (e.g. BzK or U-V vs V-K) can be applied to the parent sample of galaxies (e.g. a K-selected parent sample), so that only star forming objects are included \citep{Daddi2007,Guo2013}. In addition, it is typical to separate galaxies between red-dead and blue-star forming galaxies. \citet{Elbaz07} used the rest-frame colour-magnitude diagram (g-band  centered at 4825 $\AA$) to separate between the two populations and excluded the red objects. This selection is used to reject possible contaminations from emission line objects whose emission is not due to star-formation and typically excludes passive or highly star forming systems with large contents of dust. In contrast to the above methods, stellar mass selection includes, besides Star Forming Galaxies (SFGs), objects with high contents of dust and/or high stellar masses \citep{Kajisawa2010,Karim2011,Bauer11}. \\
The first question that arises is: Do samples of galaxies that were obtained using different selection criteria and wavelengths produce SFR$-{\rm M}_{\star}$ relations that provide similar constrains for models and theory? A comparison between different redshift results ($z \sim4-7$) would suggest that the observed SFR$-{\rm M}_{\star}$ relation could be significantly affected by the technique used to sample the  galaxies \citep{Katsianis2014}. 

In addition, different authors use different methods to obtain the intrinsic SFRs and dust corrections of the observed galaxies. The most common techniques are: \\

\noindent
1) Conversion of IR+UV luminosities to SFRs; \\ 
2) SED fitting \citep[e.g.][]{Bruzualch03} to a range of wavelengths to obtain dust corrections and/or SFRs. \\
\\
Both techniques have their advantages and shortcomings. However, the second question that arises is: Do the two above different methods produce SFR$-{\rm M}_{\star}$ relations that are consistent with each other? Interestingly, \citet{Kajisawa2010} and \citet{Bauer11} demonstrated that the above different techniques produce different SFR$-{\rm M}_{\star}$ relations for the same sample of galaxies at high redshifts. When the authors used the conversion of IR and UV luminosities, instead of SED fitting, to estimate dust corrections, they obtained higher values of SFR at a fixed stellar mass. In addition, \citet{Utomo2014} claimed that the UV and IR luminosities overestimate SFRs – compared to the SED SFRs – by more than 1 dex for galaxies with Specific SFR (sSFR $\equiv$ SFR/${\rm M}_{\star}$),  $ {\rm \log(SSFR)} < -10 \, yr^{-1}$. However, for the young highest star-forming galaxies in their sample the two methods to derive the SFRs were found broadly consistent. Similar results were supported by \citet{Fumagalli2014}. \citet{Boquien2014} argued that SFRs obtained from modeling that takes into account only FUV and U bands are overestimated. Finally, \citet{Hayward2014} noted that the SFRs obtained from IR luminosities \citep[e.g. ][]{Noeske2007,Daddi2007} can be artificially high. Despite the fact that SED fitting techniques imply SFR$-{\rm M}_{\star}$ relations that are inconsistent with results that rely solely on IR+UV-SFR conversions, it is quite common for compilation studies to combine SFR$({\rm M}_{\star})$ that were obtained employing different methods \citep[e.g. ][]{Behroozi2013}. In Table \ref{Observationspre} we present a summary of the different observations used for this work. We include the technique, sample selection, mass limit and area of sky covered. We will make a brief comparison between these to investigate if there is a tension between the relations reported by different authors at $z \sim1-4$. However, we will mostly focus on the SFR$-{\rm M}_{\star}$ relations that simulations produce and how different they are from observations.

\begin{table*}[!t]
\centering
\caption{Summary of the different observations used for this work.}
\resizebox{0.95\textwidth}{!}{%
\begin{tabular}{llccccccccccc}
  \\ \hline & Publication & Redshift & Technique to obtain  & Parent sample & Further selection-  &  Original IMF &  \\ & & &  SFR-dust corrections & Wavelength/technique & Final sample  &  \\ \hline \hline

& \citet{Noeske2007} & 0.8, 1.0 & EL+IR  & Optical  & Main Sequence-SFG  & \citet{kroupa01} & \\ \hline

& \citet{Elbaz07} & 1.0 & UV+IR  & UV+optical  & Blue-SFG & \citet{kroupa01} & \\ \hline

& \citet{Daddi2007}$^{1}$ & 2.0 & UV+IR  & K-band &  BzK-SFG    & \citet{salpeter55}  & \\ \hline

& \citet{Drory08} & 2.6-3.8 & SED(2800$\AA$)-$A_{SED}$ $^a$  & I-band  & None-Mag limited & \citet{salpeter55} & \\ \hline

& \citet{Magdis10} & 3.0 & UV+IR  & Optical  & LBG & \citet{salpeter55} & \\ \hline

& \citet{Kajisawa2010} & 0.75-3.0 & SED(2800$\AA$)-E(B-V)$_{SED}^b$+IR  & K-band & Mass  &  \citet{salpeter55} & \\ \hline

& \citet{Kajisawa2010}$^2$ & 0.75-3.0 & IR+UV  & K-band & Mass  & \citet{salpeter55} & \\ \hline

& \citet{Karim2011} & 0.8-3 & Radio  & Optical  & Mass & \citet{Chabrier03} & \\ \hline

& \citet{Bauer11} & 1.75-2.75 & SED(2800 $\AA$)-$UVslope_{SED}$$^c$  & Multi-wavelength  & Mass &  \citet{salpeter55} & \\ \hline

& \citet{Bauer11}$^2$ & 1.75-2.75 & IR+UV  & Multi-wavelength & only 24 $\mu m$ detected  &  \citet{salpeter55} & \\ \hline

& \citet{reddy2012} & 2.0 & UV+IR  & UV-LBG & LBG, Malmquist bias   & \citet{salpeter55} & \\ \hline

& \citet{reddy2012}$^3$ & 2.0 & UV+IR  & UV-LBG  & LBG, bias corrected  & \citet{salpeter55} & \\ \hline

& \citet{bouwens2012} & 3.8 & UV-UVslope & UV-LBG & LBG & \citet{salpeter55} & \\ \hline

& \citet{Kashino2013} & 1.5 & H$\alpha$+UV & Optical & Bzk-SFG & \citet{salpeter55} & \\ \hline

& \citet{Behroozi2013} & 0-4 & Comp & Comp &  Comp & \citet{Chabrier03} & \\ \hline

& \citet{Bauer2013} & Local & H$\alpha$-Balmer decrement  & Multi-wavelength &  Mass & \citet{Chabrier03} & \\ \hline

& \citet{Koyama2013} & 0.8, 2.2 & H$\alpha$-\citet{Garn2010}  & H$\alpha$  & H$\alpha$   & \citet{salpeter55} & \\ \hline

& \citet{Guo2013} & 0.7 & UV+IR  & Multi-wavelength  & U-V vs V-k -SFG & \citet{Chabrier03} & \\ \hline

& \citet{Heinis2014} & 1.5, 3.0, 4.0 & UV+IR  & UV-LBG  & LBG   & \citet{Chabrier03} & \\ \hline

& \citet{Whitaker2014} & 0.5-2.5 & UV+IR  & NIR & U-V vs V-J -SFG  & \citet{Chabrier03} & \\ \hline

& \citet{deBarros} & 3.0, 4.0 & SED-$A_{SED}$ $^d$   & UV-LBG & LBG & \citet{salpeter55} & \\ \hline

& \citet{Delosrayes2014} & 0.8 & H$\alpha$-$A_{SED}$ $^e$  & H$\alpha$ & H$\alpha$  & \citet{Chabrier03} & \\ \hline

& \citet{Salmon2015} & 4.0 & SED-$A_{SED}$ $^f$  & Multi-wavelength & None & \citet{salpeter55} & \\ \hline

& \citet{Tomczak2016} & 0.8-4.0 & UV+IR & Multi-wavelength & Mass & \citet{Chabrier03} & \\ \hline  
\end{tabular}%
}
\vspace{-0.1cm}
\medskip\\
\begin{tablenotes}
\item {\bf Notes:} Column 1, publication reference of the observed SFR$-{\rm M}_{\star}$ relation; column 2,
  redshift used; column 3, technique and type of luminosity used to obtain the intrinsic SFR and dust corrections. These are: Emission lines (EL), Infrared (IR) luminosity, Ultra-Violet (UV) luminosity, Spectral Energy Distribution (SED) and radio luminosity. We include the Compilation (Comp) of studies presented in \citet{Behroozi2013};  column 4, parent sample selection method/wavelength; column 5, final sample selection method. These are: Star Formation Galaxy (SFG),  mass, H$\alpha$ and Lyman Break Galaxy (LBG) selection; column 6, mass limit of the observations; column 7, area covered by the survey in $deg^2$; column 8, original IMF adopted. We present the relations suggested by the authors corrected to a \citet{Chabrier03} IMF when necessary (this conversion does not significantly affect the relations given since a similar change to SFR and ${\rm M}_{\star}$ is used) along with results from cosmological hydrodynamic simulations in  Figure  \ref{fig:SFR-mass42} ($z \sim 3.1$ and $z \sim 3.8$), Figure  \ref{fig:SFR-mass422} ($z \sim 2.2$ and $z \sim 2.6$), Figure  \ref{fig:SFR-mass42211} ($z \sim 2.0$ and $z \sim 1.5$) and Figure  \ref{fig:SFR-mass4222223} ($z \sim0.8-1.15$). 1) \citet{Daddi2007} used a fraction of the GOODS-N and GOODS-S fields, but did not report an exact area. 2) These observations are for the same sample of galaxies of the original work but adopt different methods for the determination of intrinsic SFRs and dust corrections. 3) This set of observations is corrected for the effects of the Malmquist bias. \\ a) SFRs from rest-frame UV (2800 $\AA$) which was calculated from the galaxy SEDs (U, B, g, R, I, 834 nm, z, j and K bands). Dust corrections using the extinction curve of \citet{Calzetti1997}.\\b) SFRs from rest-frame UV (2800 $\AA$) which was calculated from SEDs using multi-band photometry (U, B, V, i, z, J, H, K, 3.6 $\mu$m, 4.5 $\mu$m, and 5.8 $\mu$m). \\c) SFRs from Optical ACS/$z_{850}$-band. Dust corrections from UV slope calculated from multiwavelength SED-fitting.  \\d) SFRs from SEDs (B, V, I, z, U, R, J, H and K bands)+the effect of nebular emission lines. Dust corrections using the extinction curve of \citet{Calzetti2000}. \\e) SFR from H$\alpha$ luminosity. Dust corrections from the SEDs of galaxies at  rest-frame UV and optical bands. \\f) SFRs from SEDs (B, V, i, I, z, Y, J, JH and H bands). Dust corrections using \citet{Pei1992} and the extinction curve of \citet{Calzetti2000}. 

\end{tablenotes}
\label{Observationspre}
\end{table*}

Motivated by its importance in understanding galaxy evolution, a number of authors \citep[e.g.][]{Dave08,Dutton10,Finlator11,DayalFerrara2012,Kannan2014,Sparre2014,Furlong2014,Katsianis2014} have used hydrodynamic and semi-analytic models to predict the SFR$-{\rm M}_{\star}$ relation and its evolution. Numerical results \citep[e.g.][]{Dave08} predict a steeper relation than is found in observations. For example, the SFR$-{\rm M}_{\star}$ relation presented by \citet{Kannan2014} provides good agreement with the observed relation of \citet{Kajisawa2010} for $z \sim3$. On the other hand, for redshift $z \sim2.2$ the simulated galaxies have a star formation rate that is only half as large as observed. \citet{Katsianis2014} reported a discrepancy between the simulated SFR$-{\rm M}_{\star}$ relation and the observed relations of \citet{bouwens2012} and \citet{Heinis2014} at $z \sim4$. On the other hand, good agreement was found between numerical results and the observations of \citet{Drory08} and \citet{deBarros}.  More recently, \citet{Sparre2014} used  high resolution simulations to investigate the SFR$-{\rm M}_{\star}$ relation for redshifts $z \sim0-4$ as part of the Illustris project \citep{Genel2014}. At $z \sim4$ and $z \sim0$, the authors are broadly in agreement with the relation given by the compilation of observations used by  \citet{Behroozi2013}. However, the  normalization of the simulated relation is significantly lower than the observational constraints at $z \sim1$ and $z \sim2$. In addition, the simulated relation is steeper than observed at all redshifts. In another study, \citet{Furlong2014} showed agreement between simulated and observed specific SFR - stellar mass relations at $z \sim0$  using high resolution simulations from the EAGLE project \citep{Schaye2015}.  However, once again, the observed relations have a significantly higher normalization at $z \sim1$ and $z \sim2$. The tension between observations and simulations,  especially at intermediate redshifts, implies either that the current models of galaxy evolution are incomplete or that observations are being misinterpreted.

This paper is the third of a series, in which we seek to study a range of models for galaxy formation in hydrodynamical simulations through comparisons with observations of SFR and stellar mass across cosmic time.  In the first paper \citep{TescariKaW2013} we constrained and compared our hydrodynamic simulations with observations of the cosmic star formation rate density and Star Formation Rate Function (SFRF) at $z \sim 4-7$. In the second paper \citep{Katsianis2014} we demonstrated that the same cosmological hydrodynamic simulations reproduce the observed Galaxy Stellar Mass Function (GSMF) and SFR$-{\rm M}_{\star}$  relation for the same redshift interval. In this work, we extend the analysis presented in \citet{Katsianis2014} down to $z \sim1$, and critically address the comparison of SFR$-{\rm M}_{\star}$ relations with observations obtained by using different analysis techniques. In section \ref{thecode} we briefly summarize our numerical methodology. In section \ref{SIMSVS} we compare our results with the simulated SFR$-{\rm M}_{\star}$ relations from different groups and demostrate the excellent consistency between numerical results from various projects (e.g. {\textsc{Angus}}, Illustris, EAGLE). In section \ref{CompaSFRSM}  we present the evolution of the SFR$-{\rm M}_{\star}$ relation alongside observations from different groups that used different techniques to obtain their results. We draw our conclusions in section \ref{Disc}. In Appendix \ref{Comparisonold} we discuss how the uncertainty of the observed SFR$-{\rm M}_{\star}$ relation has affected the comparison with simulations in the past.


\section{Simulations}
\label{thecode}

In this work we use the set of {\textit{AustraliaN {\small{GADGET-3}} early Universe Simulations}} ({\textsc{Angus}}) described in \citet{TescariKaW2013}\footnote{ The features of our code are extensively described in \citet{TescariKaW2013} and \citet{Katsianis2014}, therefore we refer the reader to those papers for additional information.}. We run these simulations using the hydrodynamic code {\small{P-GADGET3(XXL)}}. We assume a flat $\Lambda$ cold dark matter ($\Lambda$CDM) model with $\Omega_{\rm 0m}=0.272$, $\Omega_{\rm 0b}=0.0456$, $\Omega_{\rm \Lambda}=0.728$, $n_{\rm
  s}=0.963$, $H_{\rm 0}=70.4$ km s$^{-1}$ Mpc$^{-1}$ (i.e. $h = 0.704$) and $\sigma_{\rm 8}=0.809$. Our configurations have box size $L = 24$ Mpc/$h$, initial mass of the gas particles M$_{\rm GAS}=7.32\times 10^6$ M$_{\rm \odot}/h$ and a total number of particles equal to $2\times288^3$. All the simulations start at $z=60$ and were stopped at $z=0.8$. For this work we use a \citet{Chabrier03} Initial Mass Function (IMF) for all configurations.

\begin{table*}
\centering
\caption{Summary of the different runs used in this work.}
\resizebox{0.95\textwidth}{!}{%
\begin{tabular}{llccccccc}
  \\ \hline & Run & IMF & Box Size & N$_{\rm TOT}$ & M$_{\rm DM}$ & M$_{\rm GAS}$  &
  Comoving Softening & Feedback \\ & & & [Mpc/$h$] &
  & [M$_{\rm \odot}$/$h$] & [M$_{\rm \odot}$/$h$] & [kpc/$h$] \\ \hline \hline
  & \textit{Ch24\tu eA\tu CsW} & Chabrier & 24 & $2\times288^3$ &
  3.64$\times10^{7}$ & $7.32\times 10^6$ & 4.0 & Early AGN $+$ Constant strong Winds \\ \hline
  & \textit{Ch24\tu eA\tu nW} & Chabrier & 24 & $2\times288^3$ & 3.64$\times10^{7}$ & $7.32\times 10^6$ & 4.0 & Early AGN $+$ no Winds  \\ \hline
  & \textit{Ch24\tu NF} & Chabrier & 24 & $2\times288^3$ &
  3.64$\times10^{7}$ & $7.32\times 10^6$ & 4.0 & No Feedback\\ \hline
  & \textit{Ch24\tu eA\tu MDW}$^a$ & Chabrier & 24 & $2\times288^3$ &
  3.64$\times10^{7}$ & $7.32\times 10^6$ & 4.0 & Early AGN $+$ \\
  & & & & & & & & Momentum-Driven Winds \\ \hline
  & \textit{Ch24\tu eA\tu EDW}$^b$ & Chabrier & 24 & $2\times288^3$ &
  3.64$\times10^{7}$ & $7.32\times 10^6$ & 4.0 & Early AGN $+$ \\ 
  & & & & & & & & Energy-Driven Winds \\ \hline
  & \textit{Ch24\tu Zc\tu eA\tu CsW}$^c$ & Chabrier & 24 & $2\times288^3$  & 3.64$\times10^{7}$ & $7.32\times 10^6$ & 4.0 & Early AGN $+$ Constant strong Winds  \\ 
  & & & & & & & & Metal cooling \\ \hline
  & \textit{Ch24\tu Zc\tu eA\tu EDW}$^c$ & Chabrier & 24 & $2\times288^3$  & 3.64$\times10^{7}$ & $7.32\times 10^6$ & 4.0 & Early AGN $+$ Energy-Driven Winds  \\ 
  & & & & & & & & Metal cooling \\ 

  \hline \hline \\
\end{tabular}%
}
\begin{tablenotes}
\item 
{\bf Notes:} Column 1, run name; column 2,
  Initial Mass Function (IMF) chosen; column 3, box size in comoving Mpc/$h$;
  column 4, total number of particles (N$_{\rm TOT} =$ N$_{\rm
    GAS}$ $+$ N$_{\rm DM}$); column 5, mass of the dark matter particles; column 6, initial mass of the gas particles;
  column 7, Plummer-equivalent comoving gravitational softening length; column 8, type of feedback
  implemented. See section \ref{thecode} and \citet{TescariKaW2013} for
  more details on the parameters used for the different feedback
  recipes. $(a)$: in this simulation we
  adopt variable momentum-driven galactic winds (Subection \ref{Feed}).
 $(b)$: in this simulation we
  adopt variable energy-driven galactic winds (Subsection \ref{Feed}).  $(c)$: in these simulations the effect of metal-line cooling is
  included (Subsection \ref{Zc}). For all the other runs we use cooling tables for gas of primordial composition (H + He).
\end{tablenotes}
\label{tab:sim_runs}
\end{table*}

We explore different feedback prescriptions, in order to understand the origin of the difference between observed and simulated relationships. We also study the effects of metal cooling.  We do not explore the broadest possible range of simulations, but concentrate on the simulations that can describe the high-$z$ star formation rate function and galaxy stellar mass function. We performed resolution tests for high redshifts ($z \sim4-7$) in the appendix of \citet{Katsianis2014} and showed that our results converge for objects with $\log_{10} \le ({\rm M}{_\star}/{\rm M}_{\odot}) \ge 8.5$. In Table \ref{tab:sim_runs} we summarise the main parameters of the cosmological simulations performed for this work.

\subsection{SNe feedback}
\label{Feed}

We investigate the effect of three different galactic winds schemes in the simulated SFR$-{\rm M}_{\star}$ relation. We use the implementation of galactic winds of \citet{springel2003}. We assume the wind mass loading factor  $\eta=\dot{\rm M}_{\rm w}/\dot{\rm M}_{\rm \star}=2$ and a fixed wind velocity $v_{\rm w}=450$ km/s. In addition, we explore the effects of variable winds. We use a momentum driven wind model in which the velocity of the winds is proportional to the circular velocity $v_{\rm circ}$ of the galaxy:
\begin{eqnarray}
  v_{\rm w}= 2\;\sqrt{\frac{G{\rm M}_{\rm halo}}{R_{\rm 200}}}=2\times v_{\rm circ},
\end{eqnarray}
and the loading factor $\eta$,
\begin{eqnarray}
  \eta = 2\times\frac{450\,\,{\rm km/s}}{v_{\rm w}},
\end{eqnarray}
where  M$_{\rm halo}$ is the halo mass and $R_{\rm 200}$ is the radius within which a density 200 times the mean density of the Universe at redshift $z$ is enclosed \citep{barai13}. Furthermore, we investigate the effect of the energy driven winds used by \citet{PuchweinSpri12}. In this case the loading factor is
\begin{eqnarray}
  \eta = 2\times\left(\frac{450\,\,{\rm km/s}}{v_{\rm w}}\right)^2,
\end{eqnarray}
while $v_{\rm w} = 2\times v_{\rm circ}$.

\subsection{AGN feedback}
\label{AGN}

In our scheme for Active Galactic Nuclei (AGN) feedback, when a dark matter halo reaches a mass above a given mass threshold  M$_{\rm th}=2.9\times10^{10}$ M$_{\rm\odot}/h$  for the first time, it is seeded with a central Super-Massive Black Hole (SMBH) of mass M$_{\rm seed}=5.8\times 10^{4}$  M$_{\rm\odot}/h$ (provided it contains a minimum mass fraction in stars $f_{\star}=2.0\times10^{-4}$). Each SMBH will then increase its mass by accreting local gas from a maximum accretion radius $R_{\rm ac}=200$ kpc/$h$. In this scheme we allow the presence of a black hole in low mass halos, and at early times. The AGN feedback prescription that we use combined with efficient winds is successful at reproducing the observed SFRF \citep{TescariKaW2013} and GSMF \citep{Katsianis2014} for redshifts $ 4 < z < 7$. 
 
\subsection{Metal cooling}
\label{Zc}

Our code follows the evolution of 11 elements (H, He, C, Ca, O, Ne, Mg, S, Si and Fe) released from supernovae (SNIa and SNII) and low and intermediate mass stars self-consistently \citep{T07}. Radiative heating and cooling processes are included  following \citet{wiersma09}. We assume a mean background radiation composed of the cosmic microwave background and the \citet{haardtmadau01} ultraviolet/X-ray background from quasars and galaxies. Contributions to cooling from each one of the eleven elements mentioned above have been pre-computed using the {\small Cloudy} photo-ionisation code \citep[last described in][]{ferland13} for an optically thin gas in (photo)ionisation equilibrium. In this work we use cooling tables for gas of primordial composition (H + He) as the reference configuration. To test the effect of metal-line cooling, we include it in two simulations (\textit{Ch24\tu Zc\tu eA\tu EDW} and \textit{Ch24\tu Zc\tu eA\tu CsW}).

\section{Comparison between SFR$-{ \bf \rm M}_{\star}$ relations from different simulations}
\label{SIMSVS}

Cosmological hydrodynamic simulations provide a powerful tool to investigate and  predict properties of galaxies and their distribution. Recently, the Illustris \citep{Sparre2014} and EAGLE \citep{Furlong2014} projects have used high resolution simulations and tried to reproduce the observed SFR$-{\rm M}_{\star}$. In addition, semi-analytic models have been used to reproduce the observed relations at various redshifts \citep{Dutton10}. The main questions that arise are: \\

 \begin{figure*}[!t] 
\centering
\vspace{0.0cm}
\hspace{0.0cm}
\includegraphics[scale=0.8]{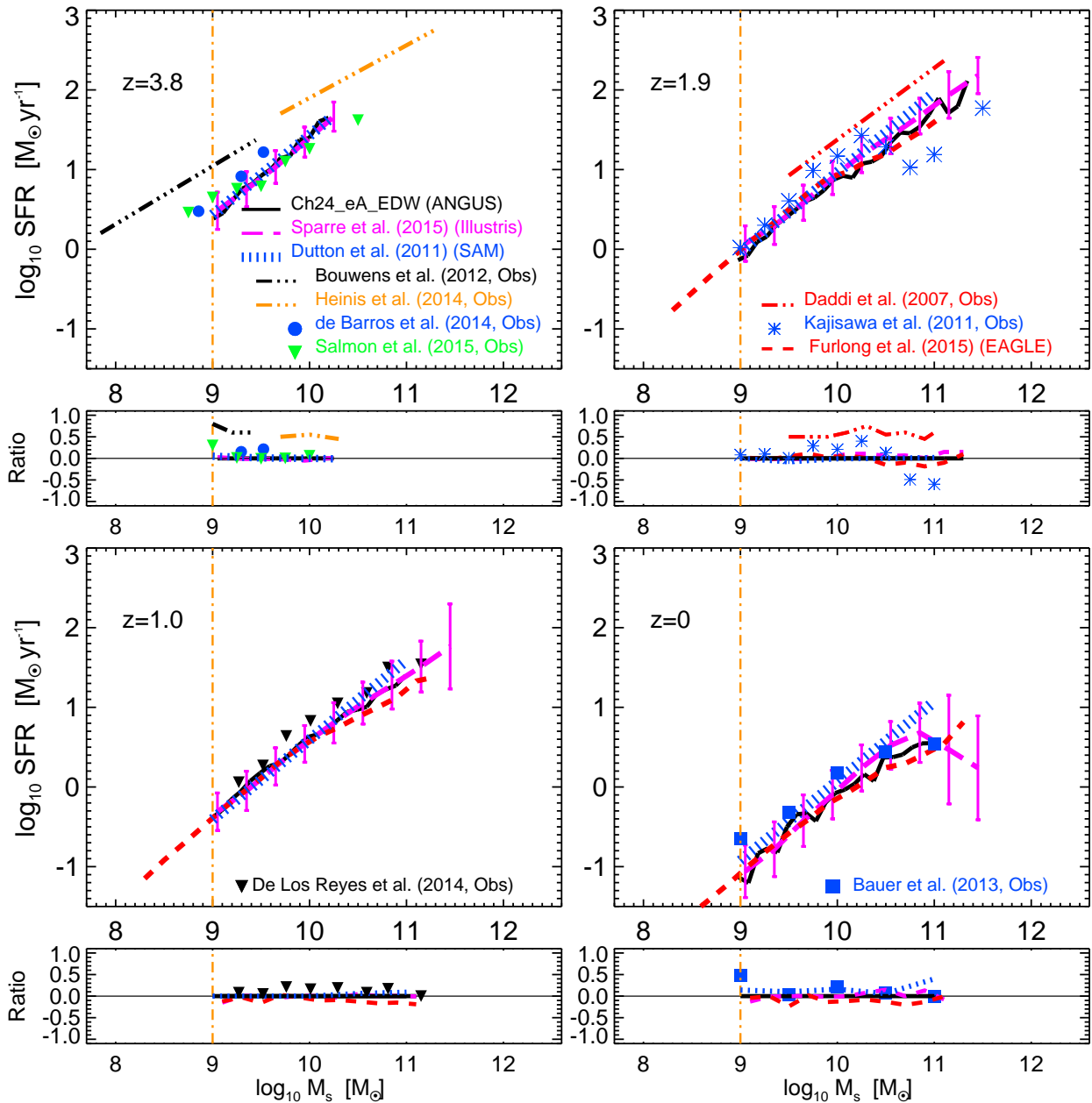}
\vspace{0.32cm}
\caption{Median values of the SFR$-{\rm M}_{\star}$ relations from different cosmological hydrodynamic simulations for $z \sim 0 - 4$. The black line is the median line through all the points of the scatter plot for our reference model (\textit{Ch24\tu eA\tu EDW}). The blue dotted line is the median fit of the scatter plot obtained with the Semi-Analytic Model (SAM) of \citet{Dutton10}. The magenta dashed line is the median line of the scatter plot presented by \citet[][Illustris project]{Sparre2014}. The red dashed line is the median line of the scatter plot presented by \citet[][EAGLE project]{Furlong2014}. We cut our ${\rm SFR}({\rm M}_{\star})$  under our confidence mass limit of $10^{9}$ M$_{\rm \odot}$ to make a meaningful comparison with the Illustris and EAGLE projects. There is an excellent agreement between the results from cosmological hydrodynamic simulations run by different groups. At each redshift, a panel showing ratios between the different simulations and observations with the \textit{Ch24\tu eA\tu EDW} (black solid line) is included.}
\label{fig:CompSFR}
\end{figure*}

\begin{table*}
\centering
\caption{Summary of the different simulated SFR$-{\rm M}_{\star}$ relations used for Figure \ref{fig:CompSFR}.}
\resizebox{0.98\textwidth}{!}{%
\begin{tabular}{llccccccc}
  \\ \hline & Publication & redshift & Box Size & N$_{\rm TOT}$ & M$_{\rm DM}$ & M$_{\rm GAS}$  &
  Comoving Softening & Feedback \\ & & & [Mpc/$h$] &
  & [M$_{\rm \odot}$/$h$] & [M$_{\rm \odot}$/$h$] & [kpc/$h$] \\ \hline \hline
  & \textit{Ch24\tu eA\tu EDW} (this work) & 0, 1.15, 2.0, 3.8 & 24 & $2\times288^3$ &
  3.64$\times10^{7}$ & $7.32\times 10^6$ & 4.0 & Early AGN $+$ \\
  & & & & & & & & Energy-Driven Winds \\ \hline

  & \citet{Sparre2014} & 0, 1, 2, 4 & 75  & $2\times1820^3$  & 4.41$\times10^{6}$   & 8.87$\times10^{5}$  & 0.704  & AGN $+$ Stellar \\
& & & & & & & & \citet{Vogelsberger2013} \\ \hline

  & \citet{Furlong2014} & 0, 1, 2 & 70.4   & $2\times1504^3$  & 6.83$\times10^{6}$   & 1.27$\times10^{6}$  & 1.87  &  AGN $+$ Stellar \\
& & & & & & & & \citet{Crain2015} \\ \hline

& \citet{Dutton10} & 0, 1, 2, 4 & NA  & NA  & NA  & NA  & NA  & Stellar \\
& Semi-analytic & & & & & & & \citet{Dutton10} \\ \hline \hline 
\end{tabular}%
}
\vspace{0.12cm}
\begin{tablenotes}
\item {\bf Notes:} Column 1, publication reference; column 2, redshifts considered; column 3, box size of the simulation in comoving Mpc/h;  column 4, number of particles used; column 5, mass of the dark matter particles; column 6,  initial mass of the gas particles; column 7, comoving softening length; column 8, feedback prescriptions used.  The box size and masses of the dark matter and gas particles are in Mpc/$h$ and M$_{\rm \odot}$/$h$, respectively, rescaled to our adopted cosmology ($h = 0.704$).
\end{tablenotes}
\label{Simulationsprecom}
\end{table*}

\noindent
a) Are simulations and theory capable of reproducing the observed SFR$-{\rm M}_{\star}$ relations? \\ 
b) Are the results of simulations that are tuned to reproduce certain observables consistent with each other? \\
c) Cosmological hydrodynamic simulations have been evolving remarkably in the last decade in terms of resolution and box size. However, numerical modeling of the interstellar medium (ISM) and star formation physics remained essentially the same (e.g. stochastic formation of star particles in a multiphase ISM). Therefore, how different are the SFR$-{\rm M}_{\star}$ relations found by state-of-the-art simulations with respect to those found in the past? \\

In Figure \ref{fig:CompSFR} we present a comparison between the simulated median SFR$-{\rm M}_{\star}$ relations of our reference model (Project name: {\textsc{Angus}}, black solid line), and those presented in \citet[][blue dotted line, semi-analytic model]{Dutton10}, \citet[][Illustris, magenta dashed line]{Sparre2014} and \citet[][EAGLE, red dashed line]{Furlong2014}. For this analysis we do not include galaxies that have masses lower than the confidence limit of $10^{9}$ M$_{\rm \odot}$ from our simulations. This is done to make a meaningful comparison with the Illustris and EAGLE projects. We see the agreement between different groups is excellent despite the fact that they used different resolutions and box sizes (in Table \ref{Simulationsprecom} we include some details for the runs that were used to produce the simulated SFR$-{\rm M}_{\star}$ shown in Figure \ref{fig:CompSFR}). Models that are tuned to reproduce certain observables (eg. {\textsc{Angus}} - cosmic star formation rate density evolution, EAGLE - GSMF at $z \sim 0$, \citet{Dutton10} - SFR$-{\rm M}_{\star}$ relation at $z \sim 0$) produce similar results for the star formation rate main sequence. It is worth to mention that even under our mass confidence limit our results are in agreement with \citet{Furlong2014}. 

\citet{Dave08} using {\small{GADGET-2}} \citep{springel2005} demonstrated that simulations of galaxy formation produce similar SFR$-{\rm M}_{\star}$ relations, to a large extent independently of modeling details (e.g feedback prescriptions). According to the authors, this is a generic consequence of smooth and steady cold accretion which dominates the growth of the simulated objects. In addition, \citet{Dutton10} state that the SFR$-{\rm M}_{\star}$ relation is generally found to be independent of feedback, since feedback regulates the outflow rate, and mostly acts to shift galaxies along the SFR sequence, leaving the zero point of the relation invariant. We display the excellent agreement between the SFR$-{\rm M}_{\star}$ relations found in different cosmological hydrodynamic simulations ({\textsc{Angus}}, Illustris, EAGLE) and the semi-analytic results of \citet{Dutton10}. This strongly suggests that the slope and normalization of the relation have their origins in fundamental principles and assumptions commonly adopted in numerical codes, while the small differences in feedback prescriptions play a negligible role. Our results for the local Universe are consistent with the relations of other groups, despite the fact that our box size and resolution are not sufficient to robustly probe other properties of galaxies at $z \sim0$.

Increasing the resolution provides a description of smaller masses and scales. In addition, decreasing the softening length will better resolve the feedback mechanisms. For this reason, some authors claim that an exact convergence when resolution is changed is not to be expected \citep{Schaye2015}. However, \citet{Murante2015}, who used simulations of disc galaxies (based on the GADGET-3 code), demonstrated that numerical results are remarkably stable against resolution even in runs dedicated to galactic scales. For their ISM multiphase  model the authors employed differential equations that describe the evolution of a system composed of cold clouds (where stars form) embedded in hot ambient gas, at unresolved scales (MUPPI). The results from different runs are stable as resolution is decreased even by a factor of 8. In particular, morphology-related quantities, such as rotation curves and circularity histograms, vary by less than 10 per cent. The SFR varies approximately by the same amount. \citet{Murante2015} suggest that reducing the softening in simulations of disk galaxies by a factor of 6 induces effects related to numerical heating that change their morphologies and central velocities. Doubling the softening parameter results in thicker and less extended discs and increases the bulge mass. However, the results for the integrated properties are not significantly affected. The remarkable agreement between the simulated ${\rm SFR-M_{\star}}$ relations presented in Figure \ref{fig:CompSFR} suggests that cosmological simulations are able to produce realistic populations of galaxies within representative cosmological volumes, even at relatively modest resolutions. Moreover, it is a strong indication that numerical codes that rely on similar multiphase models and are tuned to reproduce statistical properties of galaxies are ``bound'' to produce similar results by construction, regardless of resolution and box size. 

In the following sections we investigate in more detail the  redshift evolution of the simulated SFR$-{\rm M}_{\star}$ relations in the {\textsc{Angus}} project and critically compare the numerical results with observations.

\section{Evolution of the SFR$-{\rm M}_{\star}$ relation from $z \sim4$ to $z \sim1$}
\label{CompaSFRSM}

\begin{figure*}[!t]
\begin{center}
\includegraphics[scale=0.7]{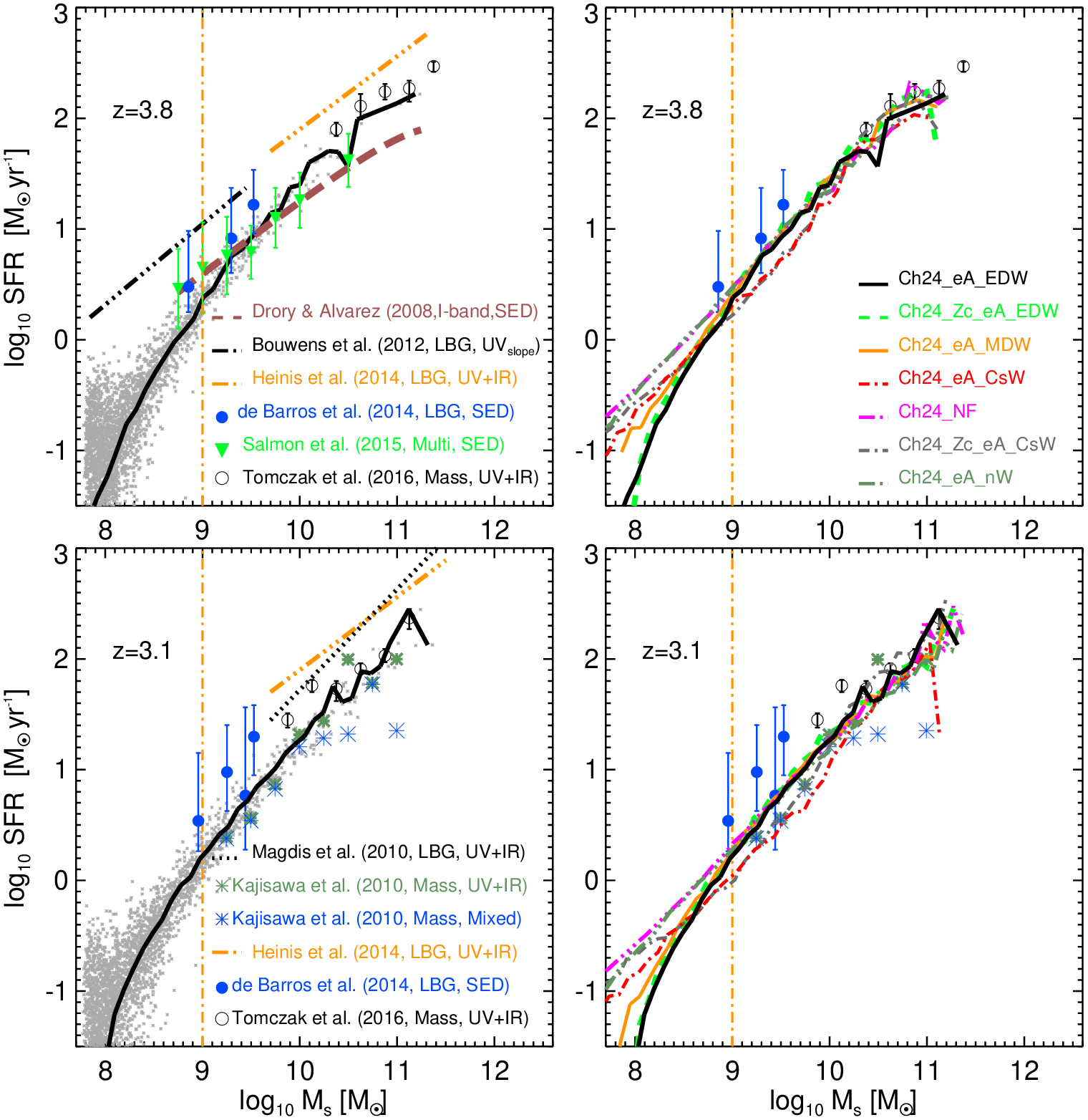}\\ 
\vspace{0.18cm}
\end{center}
\caption{Left panels: scatter plot of the SFR$-{\rm M}_{\star}$ relation for our fiducial run \textit{Ch24\tu eA\tu EDW} (grey points) at redshifts $z\sim3.1$ and $3.8$. In each panel, the black solid line is the median line through the points of the scatter plot. Overplotted are the observed galaxy SFR(M$_{\star}$) relations from \citet[][I-band selected sample, SFRs(SED) - brown dashed line]{Drory08}, \citet[][ Lyman-break selected sample, SFRs(UV+IR) - black dotted line]{Magdis10},  \citet[][mass-selected sample - the dark green stars represent SFRs that were obtained using UV+IR luminosities, while the blue stars were obtained using the SED fitting technique]{Kajisawa2010}, \citet[][Lyman-break selected sample, SFRs(UV+IRX-$\beta$) - black triple dot-dashed line]{bouwens2012}, \citet[][Lyman-break selected sample, SFRs(UV+IR) - orange triple dot-dashed line]{Heinis2014}, \citet[][Lyman-break selected sample, SFRs(SED) - blue filled circles  with error bars]{deBarros},  \citet[][multi-wavelength derived redshifts, SFRs(SED) - reverse green triangles with error bars]{Salmon2015} and \citet[][mass-selected sample, SFRs(UV+IR) -  black open circles with error bars]{Tomczak2016}. Right panels: median lines of the SFR$-{\rm M}_{\star}$ scatter plots for all the runs of Table \ref{tab:sim_runs}. In these panels we do not present the analytic expressions of the observed relations for the sake of clarity.}
\label{fig:SFR-mass42}
\end{figure*}

\begin{figure*}[!t]
\begin{center}
\includegraphics[scale=0.7]{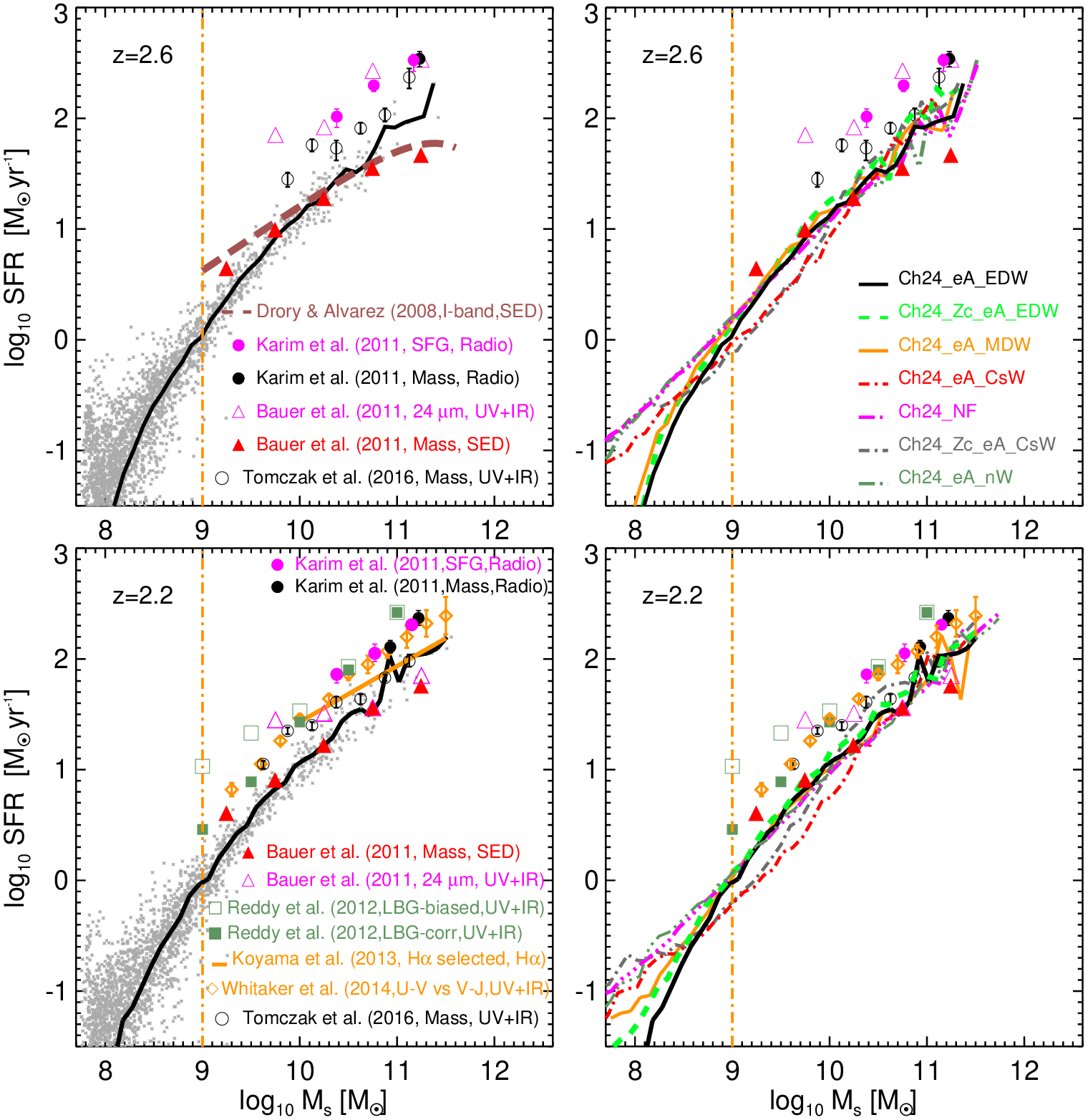}\\ 
\vspace{0.18cm}
\end{center}
\caption{Left panels: scatter plot of the SFR$-{\rm M}_{\star}$ relation for our fiducial run \textit{Ch24\tu eA\tu EDW} (grey points) at redshifts $z\sim2.2$ and $2.6$. In each panel, the black solid line is the median line through the points of the scatter plot. Overplotted are the observed galaxy SFR(M$_{\star}$) relations from \citet[][I-band selected sample, SFRs(SED) - brown dashed line]{Drory08}, \citet[][ SFG/mass-selected sample, SFRs(radio) - magenta/black circles with error bars]{Karim2011}, \citet[][mass-selected sample/SFRs that are obtained from SED fitting - red triangles]{Bauer11}, \citet[][24 $\mu$m selected sample/SFRs that are obtained from UV+IR luminosities - magenta open triangles]{Bauer11}, \citet[Lyman-break selected sample, SFRs(UV+IR) - dark green open squares are not corrected for incompleteness, dark green filled squares are the corrected results]{reddy2012}, \citet[][H$\alpha$-selected sample, SFRs(H$\alpha$) -  orange solid line]{Koyama2013}, \citet[][U-V vs V-J -SFG sample, SFRs(UV+IR) -  orange open diamonds]{Whitaker2014} and \citet[][mass-selected sample, SFRs(UV+IR) -  black open circles with error bars]{Tomczak2016}. Right panels: median lines of the SFR$-{\rm M}_{\star}$ scatter plots for all the runs of Table \ref{tab:sim_runs}. In these panels we do not present the analytic expressions of the observed relations for the sake of clarity.}
\label{fig:SFR-mass422}
\end{figure*}

\begin{figure*}[!t]
\begin{center}
\includegraphics[scale=0.7]{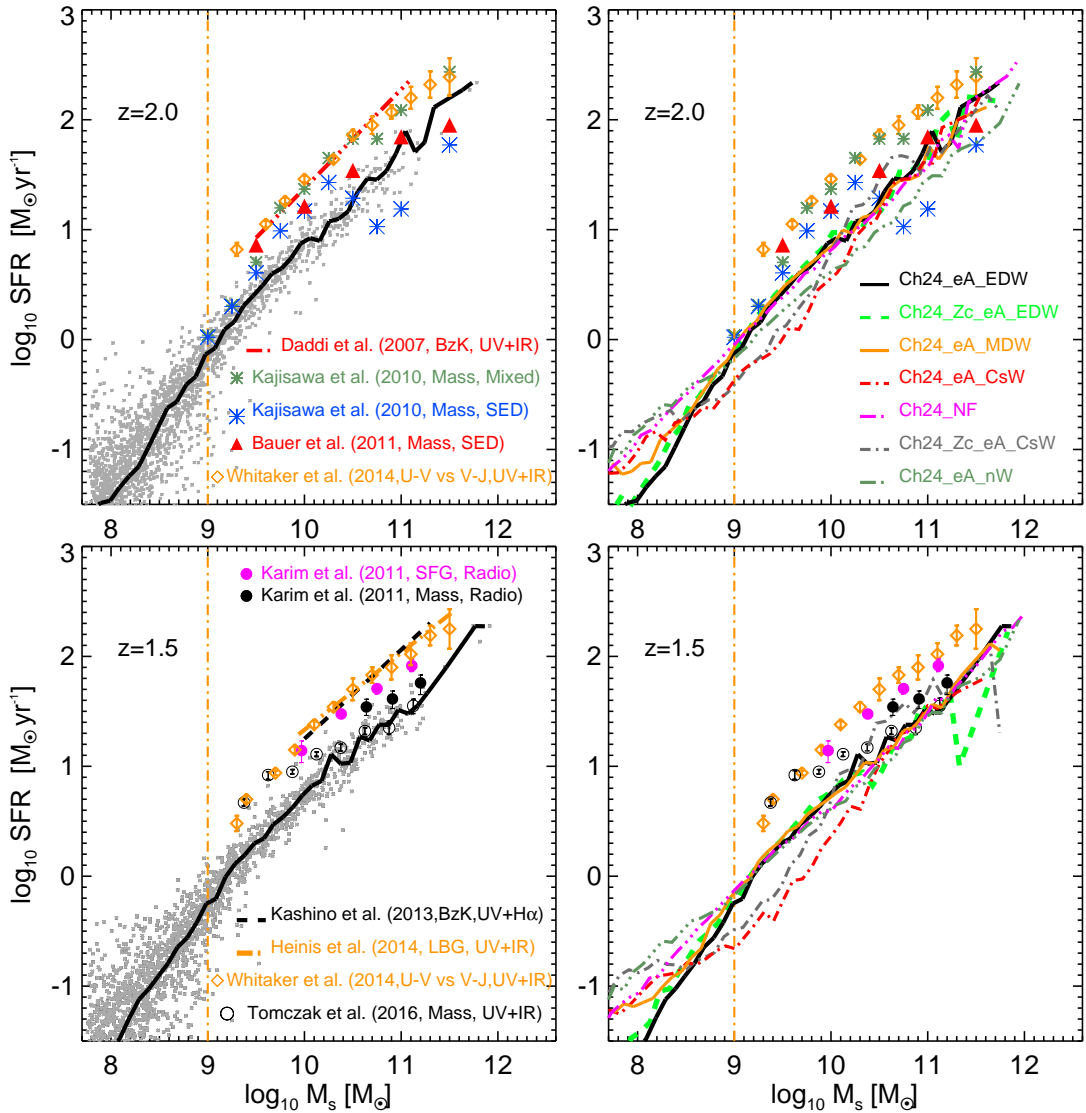}\\ 
\vspace{0.18cm}
\end{center}
\caption{ Left panels: scatter plot of the SFR$-{\rm M}_{\star}$ relation for our fiducial run \textit{Ch24\tu eA\tu EDW} (grey points) at redshifts $z\sim1.5$ and $2.0$. In each panel, the black solid line is the median line through the scatter plot. Overplotted are the observed galaxy SFR(M$_{\star}$) relations from \citet[][BzK-SFGs, SFRs(UV+IR) - red triple dot-dashed line]{Daddi2007}, \citet[][mass-selected sample, SED/UV+IR - blue stars/green stars]{Kajisawa2010}, \citet[][SFGs/mass-selected sample - magenta/black circles]{Karim2011}, \citet[][mass-selected sample, SFRs(SED) - red triangles]{Bauer11}, \citet[][Lyman-break selected sample, SFRs(UV+H$\alpha$) - black dashed line]{Kashino2013}, \citet[][Lyman-break selected sample, SFRs(UV+IR) - orange triple dot-dashed line]{Heinis2014}, \citet[][U-V vs V-J -SFG sample, SFRs(UV+IR) -  orange open diamonds]{Whitaker2014} and \citet[][mass-selected sample, SFRs(UV+IR) -  black open circles with error bars]{Tomczak2016}. Right panels: median lines of the star SFR$-{\rm M}_{\star}$ scatter plots for all the runs of Table \ref{tab:sim_runs}. In these panels we do not present the analytic expressions of the observed SFR$-{\rm M}_{\star}$ relations for the sake of clarity.}
\label{fig:SFR-mass42211}
\end{figure*}

\begin{figure*}[!t]
\begin{center}
\includegraphics[scale=0.7]{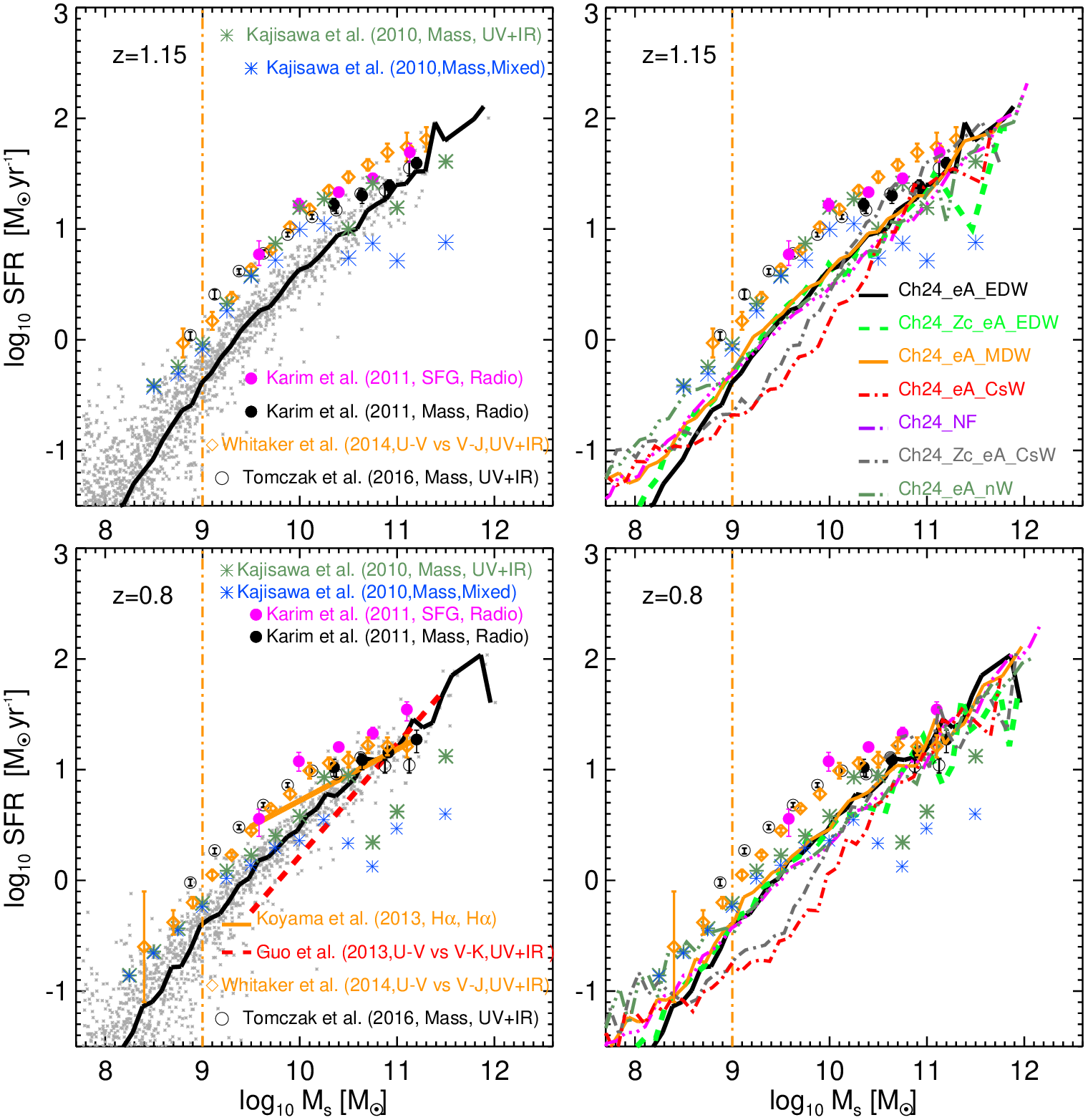}\\ 
\vspace{0.18cm}
\end{center}
\caption{Left panels: scatter plot of the SFR$-{\rm M}_{\star}$ relation for our fiducial run \textit{Ch24\tu eA\tu EDW} (grey points) at redshifts $z\sim0.8$ and $1.15$. In each panel, the black solid line is the median line through the scatter plot. Overplotted are the observed galaxy SFR(M$_{\star}$) relations from \citet[][mass-selected sample - SED/blue stars, IR+UV/dark green stars]{Kajisawa2010}, \citet[][SFGs/mass-selected sample - magenta/black circles]{Karim2011}, \citet[][U-V vs V-K SFGs, SFRs(UV+IR) -  red dashed line]{Guo2013}, \citet[][H$\alpha$-selected sample, SFRs(H$\alpha$) -  orange solid line]{Koyama2013}, \citet[][U-V vs V-J -SFG sample,  SFRs(UV+IR) -  orange open diamonds]{Whitaker2014} and \citet[][mass-selected sample, SFRs(UV+IR) -  black open circles with error bars]{Tomczak2016}. Right panels: median lines of the SFR$-{\rm M}_{\star}$ scatter plots for all the runs of Table \ref{tab:sim_runs}. In these panels we do not present the analytic expressions of the observed SFR$-{\rm M}_{\star}$ relations for the sake of clarity.}
\label{fig:SFR-mass4222223}
\end{figure*}

In \citet{Katsianis2014} we demonstrated that different observations of the SFR$-{\rm M}_{\star}$ relation are in tension for $z \sim 4-7$. This discrepancy was attributed to the different selection methods and techniques for the determination of the intrinsic SFRs and dust corrections of the observed galaxies. In this section we extend that work to $z \sim1-4$, in order to investigate if observations in this redshift interval are also in tension, and address in more detail how sample selection and methodology can affect the comparison with cosmological simulations. We present the SFR$-{\rm M}_{\star}$ relation obtained from our hydrodynamic simulations along with observations from different groups in Figs. \ref{fig:SFR-mass42} ($z \sim3.1-3.8 $), \ref{fig:SFR-mass422} ($z \sim2.2-2.6$), \ref{fig:SFR-mass42211} ($z \sim1.5-2.0$) and  \ref{fig:SFR-mass4222223} ($z \sim0.8-1.15$). In the left panels of each figure we show the scatter plots of the SFR$-{\rm M}_{\star}$ relation for our reference model \textit{Ch24\tu eA\tu EDW} (grey points). The black line is the median line through all points of the scatter plot. Our reference model, which combines a Chabrier IMF, early AGN feedback and variable energy driven winds, is able to reproduce the observed galaxy stellar mass function, star formation rate function and cosmic star formation rate density for $z \sim1-7$ galaxies \citep{TescariKaW2013,Katsianis2014}. In the right panels we compare the median lines of the scatter plots for all the runs presented in Table \ref{tab:sim_runs}. The orange vertical line at $10^{9}$ M$_{\rm \odot}$ is the confidence limit of our simulations and the mass limit of most observations for the redshifts considered in this work.

The scatter plot of the simulated SFR$-{\rm M}_{\star}$ for our reference model at $z \sim3.8$ (top left panel of Figure \ref{fig:SFR-mass42}) is consistent with the results of the I-band selected sample of \citet{Drory08}, the Lyman-Break selected sample of \citet{deBarros} and the multi-wavelength results of \citet{Salmon2015}. On the contrary, the SFR$-{\rm M}_{\star}$ relations obtained by \citet{bouwens2012} and \citet{Heinis2014}, who used the Lyman-Break technique, imply $\sim$ 3-5 times higher SFRs at a fixed stellar mass. All the above authors use various methods to obtain the intrinsic SFRs and dust attenuation effects. \citet{Drory08} and \citet{deBarros} used SED fitting techniques. On the other hand,  \citet{bouwens2012} and \citet{Heinis2014} estimated the dust attenuation effects and  SFRs in their sample using the IRX-$\beta$ relation \citep{meurer1999}, stacking techniques and the \citet{kennicutt1998} relation. The tension between the above results may be due to the fact that the authors used different methods to obtain the intrinsic properties of galaxies and/or due to selection methods. The SFR$-{\rm M}_{\star}$ given by \citet{Tomczak2016} represents a multi-wavelength sample of galaxies, where the SFRs were obtained from converting UV+IR luminosities. The relation implies lower values of SFR at a fixed stellar mass than the Lyman break selected sample of \citet{Heinis2014} who used UV+IR-SFR conversions as well. \citet{Heinis2014} discussed the possible impact of the Lyman-break selection on their retrieved SFR$-{\rm M}_{\star}$ relation, and state that their sample may not include a significant number of objects since they preferably select star forming objects. They point out that their shallow slope, $\sim 0.7$, might be also caused by the fact that they are selecting galaxies by their UV flux, which does not pick objects with low star formation rates at fixed stellar mass or dusty massive objects. The above would suggest that the differences between the results of the two authors could be attributed to the differences of their selection methods. On the other hand, the results of \citet{Tomczak2016} suggest a higher normalization than the observations of \citet{Salmon2015}. Both authors used multi-wavelength samples of galaxies but \citet{Salmon2015} used an SED method to obtain SFRs instead of a UV+IR conversion. As discussed in the introduction it is quite possible that different techniques produce different result. Differences between \citet{Salmon2015} and \citet{Tomczak2016} maybe can be attributed to the technique used to obtain dust corrections and SFRs. For redshift $z=3.8$, our configurations that combine different feedback schemes  resemble each other closely (top right panel of Figure \ref{fig:SFR-mass42}). This follows the results of \citet{Katsianis2014}, who showed that different feedback prescriptions result in roughly the same SFR$-{\rm M}_{\star}$ relation for $z \sim4-7$.

At $z \sim3.1$, our reference model (\textit{Ch24\tu eA\tu EDW}, bottom left panel of Figure \ref{fig:SFR-mass42}) is consistent with the the mass selected sample of \citet{Kajisawa2010}, and the Lyman-break selected sample of \citet{deBarros}. \citet{Kajisawa2010} used two different methods to obtain the intrinsic SFRs at a fixed stellar mass for the same sample of galaxies. In the first case, they used the sum of IR and UV light to estimate the observed SFRs for objects that were detected having an IR 24 $\mu m$ flux (originating from dusty high star forming galaxies). For the other objects (undetected at 24 $\mu m$) the authors used SED fitting techniques and UV luminosities. The dark green stars of Figure \ref{fig:SFR-mass42} are obtained with this methodology. In the second case, SED fitting technique and UV luminosities were used for the whole sample of galaxies. They find good agreement between both methods at $z \sim3.0$, even though the SFRs from the combination of UV and IR light are higher than those found by SED fitting. \citet{Kajisawa2010} stressed that, at $z \sim3$, the ratio $\log({\rm SFR}_{IR+UV}/{\rm SFR}_{UV_{SED})}$ for galaxies with 24 $\mu m$ flux is as high as $\sim 0.63$. This is an example of how  methodology can affect the determination of SFRs in a sample of galaxies. The relations given by the Lyman-break selection of \citet{Magdis10} and \citet{Heinis2014} have significantly higher normalizations than those found by SED measurements, mass selected samples and cosmological simulations.

We see the effect of  metal cooling and different feedback prescriptions among the different simulations at $z \sim3.1$ in the bottom right panel of Figure \ref{fig:SFR-mass42}.  The \textit{Ch24\tu NF} (no feedback) and \textit{Ch24\tu eA\tu nW} (early AGN, no winds) runs are almost identical. This means that the effect of our AGN feedback model on the simulated SFR$-{\rm M}_{\star}$ relation is small. Furthermore, we can compare the \textit{Ch24\tu eA\tu nW} and \textit{Ch24\tu eA\tu CsW} (early AGN, constant strong winds) to gain insight into how the constant energy wind model of \citet{springel2003} affects the simulated relation. We see that the star formation rate at fixed mass for objects with stellar mass $\log_{10} ({\rm M}{_\star}/{\rm M}_{\odot}) \le 10.0$ is lower for the \textit{Ch24\tu eA\tu CsW} run. Above the mass limit of $10^{9}$ M$_{\rm \odot}$, the \textit{Ch24\tu eA\tu MDW} (early AGN, momentum-driven winds) and \textit{Ch24\tu eA\tu EDW} models are consistent with the configurations that have no winds. This means that the effect of our variable wind models do not change the slope of the SFR$-{\rm M}_{\star}$ relation. The only small difference between our runs is found for objects with  $\log_{10} ({\rm M}{_\star}/{\rm M}_{\odot}) \le 8.5$, where there are no observations to constrain the results and feedback is not well resolved. Note that \citet{Dave08} and \citet{Dutton10} also suggest that runs with different feedback prescriptions results in similar {\rm SFR}$-{\rm M}_{\star}$ relations\footnote{Extreme feedback recipes can shape the SFR$-{\rm M}_{\star}$ relation \citep{Haas2013}. However, these runs are unable to produce galaxies with realistic SFRs and stellar masses.}. For this work we use a set of physically plausible cases that can produce realistic star formation rate and stellar mass functions in our simulations \citep{TescariKaW2013,Katsianis2014}. In addition, by comparing the \textit{Ch24\tu Zc\tu eA\tu CsW} and \textit{Ch24\tu eA\tu CsW} we see that metal cooling does not significantly change the simulated SFR$-{\rm M}_{\star}$ relation. This is due to the fact that when metal cooling is included gas can cool more efficiently. As a result the SFR increases and, correspondingly, the stellar mass increases moving galaxies along the SFR$-{\rm M}_{\star}$ relation without affecting it considerably.

The scatter plot of the simulated SFR$-{\rm M}_{\star}$ relation for our fiducial model at  $z \sim2.6$ (top left panel of Figure \ref{fig:SFR-mass422}) is consistent with the results of \citet{Drory08} and \citet{Bauer11}. The open magenta triangles show the median SFR that relied on adding IR and UV luminosities (${\rm SFR}_{IR+UV}$) for the sample of galaxies of \citet{Bauer11}, which were detected only at 24 $\mu m$. The authors state that, at $z > 2.5$, the ${\rm SFR}_{IR+UV}$ is greater than the SFR obtained from SED and UV light by an average factor of 5. The full mass selected sample of \citet{Bauer11} with dust correction laws that rely on SED fitting is almost $0.7$ dex lower and in excellent agreement with numerical results (filled red triangles). This comparison points out how selection and dust correction methods can affect the relation reported by observers at $z \sim2.6$. \citet{Bauer11} noted that the best way to robustly determine the SFR and the amount of dust extinction for each galaxy is to calculate the ultraviolet slope SED-fitting. We find that cosmological hydrodynamic simulations from various groups have a good consistency with these SED measurements of the SFR$-{\rm M}_{\star}$ relation. On the contrary, simulations are unable to reproduce the results of \citet{Karim2011}, shown by the magenta filled circles (SFGs) and black filled circles (mass selected sample), who used radio luminosities to obtain the intrinsic SFRs. \citet{Speagle14} noted  that the SFRs obtained from radio luminosities are overestimated and in tension with other SFR indicators. We see that the SED observations and numerical results have SFRs at a fixed stellar mass lower by a factor of $\sim 6$.  For redshift $z=2.6$, the \textit{Ch24\tu eA\tu CsW} and \textit{Ch24\tu Zc\tu eA\tu CsW} runs underpredict the SFR at a fixed ${\rm M}_{\star}$ for objects with stellar  masses $\log_{10} ({\rm M}{_\star}/{\rm M}_{\odot}) \le 10.2$. The rest of the configurations are consistent with each other and with the mass selected observations of \citet[][ top right panel of Figure \ref{fig:SFR-mass422}]{Bauer11}.  By comparing the \textit{Ch24\tu Zc\tu eA\tu EDW} and \textit{Ch24\tu eA\tu EDW} we see that metal cooling does not affect the simulated SFR$-{\rm M}_{\star}$ relation when energy driven winds are used. This is true for all redshifts considered in this work.

The simulated SFR$-{\rm M}_{\star}$ relation for $z \sim2.2$ (bottom left panel of Figure \ref{fig:SFR-mass422}) is consistent with the I-band selected sample of \citet{Drory08} and mass selected sample of \citet{Bauer11}. The results of \citet{Bauer11} that were obtained using two different dust correction methods (open magenta triangles and red filled triangels) agree better at $z < 2.25$.  In the bottom left panel of figure \ref{fig:SFR-mass422} we also present the LBG observations of \citet{reddy2012}. We emphasize the comparison between the biased and unbiased results of \citet{reddy2012} who investigated a set of galaxies at redshifts $1.5 < z < 2.6$. The authors quantified the effects of the non detection of faint objects in their flux limited selection. We see that the effect of the Malmquist bias (preferential selection of the most luminous-SFR galaxies at a fixed stellar mass) is important for low mass galaxies, and that the slope of the biased SFR$-{\rm M}_{\star}$ relation is therefore artificially shallow. The correction makes the slope steeper with an exponent close to unity, something that is in accordance with the predictions from cosmological simulations. We note that the constrains that include the correction for the Malmquist bias are in  very good agreement with the updated observations of \citet{Whitaker2014} and \citet{Tomczak2016}, which are able to better probe low mass objects. \citet{reddy2012}, \citet{Whitaker2014} and \citet{Tomczak2016} used combinations of IR and UV luminosities to obtain the intrinsic SFRs and dust corrections at $z \sim2.2$ and maybe this is the reason why they predict larger normalizations for their SFR$-{\rm M}_{\star}$ relation with respect to SED observations and simulations. The H$\alpha$ selected sample of \citet{Koyama2013} implies higher values of SFR at a fixed stellar mass than the mass selected sample of \citet{Bauer11}, and this maybe is due to the fact that it preferably samples high star forming galaxies ( H$\alpha$ selection). Radio SFRs are almost a factor of $\sim 4$ larger than simulations estimates at $z \sim2.2$. For redshift $z=2.2$, the configurations with constant energy driven winds underpredict the SFR at a fixed mass for objects with $\log_{10} ({\rm M}{_\star}/{\rm M}_{\odot}) \le 10.3$ (bottom right panel of Figure \ref{fig:SFR-mass422}).  However, the simulation with metal cooling and constant energy driven winds slightly overpredicts the SFR at fixed stellar mass for objects with $\log_{10} ({\rm M}{_\star}/{\rm M}_{\odot}) \ge 10.5$ .

The SFR$-{\rm M}_{\star}$ relation for our reference model at $z \sim2.0$ (top left panel of Figure \ref{fig:SFR-mass42211}) is consistent with the mass selected samples of \citet[][SED]{Kajisawa2010} and \citet{Bauer11}. However, we note that the observed relations are shallower. In contrast, the results from the BzK-SFGs of \citet{Daddi2007}, the mass selected sample of \citet[][UV+IR]{Kajisawa2010} and the U-V vs V-J -SFGs of \citet{Whitaker2014} imply a significantly higher normalization than SED measurements. \citet{Bauer11} state that they find a flattened relation relative to \citet{Daddi2007}. According to \citet{Bauer11} this is either due to the fact that they are using a mass-complete sample instead of just star-forming galaxies, or the overestimation of the dust correction applied by \citet{Daddi2007} who used a combination of IR and UV luminosities instead of a SED analysis. Furthermore, \citet{Hayward2014} note that the overestimation of SFRs from IR luminosities  may have played an important role in the observed SFR$-{\rm M}_{\star}$ relations of \citet{Daddi2007} and suggest that the methodology used by the authors may have overestimated the SFRs at a fixed stellar mass. We see that the SFR selected results of \citet[][red triple dot-dashed line - IR+UV]{Daddi2007} and the mass selected sample of \citet[][dark green symbols - UV+IR]{Kajisawa2010} are in excellent agreement. This points to the direction that the BzK-SFG selection of \citet{Daddi2007} did not  considerably  affect the SFR$-{\rm M}_{\star}$ relation and the methodology for dust corrections and determinations of intrinsic SFRs is mostly responsible for the tension with the results of \citet[][blue symbols-UV$_{SED}$]{Kajisawa2010} and \citet{Bauer11}. \citet{Kajisawa2010} state that at $z \sim 2$ the ratio of ${\rm SFR_{UV+IR}}$ and ${\rm SFR_{UV_{SED}}}$ is  $\log({\rm SFR_{IR+UV}}/{\rm SFR_{UV_{SED}}})=0.37$, for the same sample of galaxies. We also see from Figure \ref{fig:SFR-mass42211} that the difference between the two methods for the determination of the intrinsic SFRs at a fixed stellar mass is $\sim 0.75$ dex for objects with $ > 10^{10.5}$ M$_{\rm \odot}$. Simulations are more consistent with SED observations at $z \sim2$, while the use of IR light predicts larger SFRs from both. Numerical results are consistent with the observations of \citet{Tomczak2016} for high mass objects $ > 10^{10.0}$ M$_{\rm \odot}$, but at lower masses the observations suggest significantly higher SFRs at a fixed stellar mass. Overall the observed relation is found to be shallower. For redshift $z=2.0$, the simulations with constant energy driven winds underpredict the SFR at a fixed mass for objects with $\log_{10} ({\rm M}{_\star}/{\rm M}_{\odot}) \le 10$ (top right panel of Figure \ref{fig:SFR-mass42211}). The simulation with metal cooling and constant energy driven winds overpredicts the SFR at fixed stellar mass for objects with $\log_{10} ({\rm M}{_\star}/{\rm M}_{\odot}) \ge 10$.

The simulated SFR$-{\rm M}_{\star}$ relation for the \textit{Ch24\tu eA\tu EDW} at $z \sim1.5$ (bottom left panel of Figure \ref{fig:SFR-mass42211}) is  steeper with lower normalization than the results of \citet{Heinis2014}. On the other hand, the normalization of the  SFR$-{\rm M}_{\star}$ relation obtained by their sample could be possibly larger due to their Lyman-break selection and/or the methodology used to recover the intrinsic properties of galaxies. We see that the results from Lyman-break and SFGs are in excellent agreement \citep{Kashino2013,Heinis2014,Whitaker2014} and this points to the direction that they select similar high star forming systems. The agreement of the simulations with the radio SFRs of \citet{Karim2011} is improved (especially for the mass selected sample).

The scatter plot of the simulated SFR$-{\rm M}_{\star}$ relation for our reference model at $z \sim1.15$ (top left panel of Figure \ref{fig:SFR-mass4222223}) is consistent with the results of the \citet[][for ${\rm M}_{\star} \le 10^{10.5} {\rm M}_{\odot}$]{Kajisawa2010} and \citet{Karim2011}. However, all the observations suggest significantly higher SFRs at a fixed stellar mass for small objects. We see though that the tension between the observed relations is much less.  The difference between \citet[][dark green symbols-UV+IR]{Kajisawa2010} and \citet[][blue symbols-UV$_{SED}$]{Kajisawa2010} is much smaller than that found at higher redshifts. The authors state that $\log({\rm SFR}_{IR+UV}/{\rm SFR}_{UV_{SED}})$ is $ \sim 0.25$ and $ \sim 0.19$ for $z \sim1.25$ and $z \sim0.75$, respectively. The difference between the two methods to recover the intrinsic SFR at a fixed stellar mass for objects with $M_{\star} > 10^{10.5}$ M$_{\rm \odot}$ is $ \sim 0.35$ dex and $\sim 0.25 $ dex for $z \sim 1.25$ and $z \sim0.75$, respectively (considerably lower than the 0.75 dex found at $z \sim2.0$). The turn over of the relation found in the mass selected sample of \citet{Kajisawa2010} and the I-band selected sample of \citet{Drory08} could be attributed to the fact that the authors include massive quiescent galaxies \citep{Drory08}. However, recent observations of SFGs that do not include quiescent objects show as well a curvature \citep{Whitaker2014}. Simulations (including {\textsc{Angus}, Illustris and EAGLE) are unable to reproduce this behaviour. It is quite possible that current mechanisms implemented in state-of-the-art cosmological simulations to decrease the SFRs of high mass objects (like e.g. AGN feedback) are not efficient enough}. Once again, we see that the run with constant energy driven winds (\textit{Ch24\tu eA\tu CsW}) underpredicts the SFR at a fixed mass (top right panel of Figure \ref{fig:SFR-mass4222223}). The simulation with constant energy driven winds and metal cooling underpredicts the SFR at a fixed stellar mass for $\log_{10} ({\rm M}{_\star}/{\rm M}_{\odot}) \le 10.4$. On the contrary, for objects with  $\log_{10} ({\rm M}{_\star}/{\rm M}_{\odot}) \ge 10.5$ the SFR is overpredicted.  The other configurations are consistent with each other.

The simulated SFR$-{\rm M}_{\star}$ relation for our fiducial model at $z \sim0.8$ (bottom left panel of Figure \ref{fig:SFR-mass4222223}) is consistent with the mass selected samples of \citet{Kajisawa2010} and \citet{Karim2011}, the H$\alpha$ selected sample of \citet{Koyama2013} and the U-V - V-k selected sample of \citet{Guo2013}. The updated relation given by \citet{Guo2013} is steeper than other estimates in the literature \citep[e.g. ][]{Noeske2007}. \citet{Guo2013} noted that taking into account low mass objects is critical for determining the slope of the SFR$-{\rm M}_{\star}$ relation and the reason why they find a steeper slope (with an exponential slope close to unity) is that they can take into account objects that previous surveys did not.  If the authors fit only their relation to the 24 $\mu m$ detected galaxies, they would get a much shallower slope in perfect agreement with previous estimates \citep[e.g. ][]{Noeske2007}. This result is consistent with \citet{reddy2012} and cosmological simulations. Numerical results are consistent with the observations of \citet{Whitaker2014} and \citet{Tomczak2016} for high mass objects ($M_{\star}>10^{10}$ M$_{\rm \odot}$) but there is a severe tension at lower masses. As for the higher redshifts considered in this work, we find that simulations with constant energy driven winds underpredict the SFR at a fixed mass for objects with stellar masses lower than $ \sim 10^{10.3}$ M$_{\rm \odot}$ (bottom right panel of Figure \ref{fig:SFR-mass4222223}).

In conclusion, we find that simulations show a good agreement with observations that rely on SED fitting  techniques for the determination of the intrinsic SFRs and dust corrections at $z>1.5$. On the contrary, numerical results are not consistent with the combination of UV and IR luminosities and produce lower star formation rates at a fixed stellar mass by almost a factor of 5. This finding confirms the results of \citet{Katsianis2014} which addressed higher redshifts ($z\sim 4-7$).  We demonstrate that the above is true for the simulated SFR$-{\rm M}_{\star}$ relations of various groups. It is important to note though that the physics assumed for numerical modeling are not yet optimized to reflect reality even in the state-of-the art simulations and thus it is currently impossible to determine which observational method produces robust results. For $z<1.5$ numerical results suggest that SFRs of high mass objects ($\log_{10} ({\rm M}{_\star}/{\rm M}_{\odot}) > 10.6$) that rely solely on UV luminosities could be underpredicted. This may be due to the fact that the dusty environment of high mass low redshift galaxies does not allow UV light to escape the galaxy and the ${\rm SFR}_{UV}$ ends up being underestimated. \citet{Kajisawa2010} also suggest that, if a galaxy has star-forming regions from which one can detect no UV light at all due to the heavy dust obscuration (this case occurs more frequently for low redshift galaxies with large masses), only UV light from relatively less-obscured regions contributes to the observed SED, and this results in the underestimation of the dust extinction. Furthermore, we see that sample selection can affect the results for the observed SFR$-{\rm M}_{\star}$ relation. The ratio of $\log({\rm SFR}_{Lyman,H\alpha,blue\, \, Selection}/{\rm SFR}_{Mass \, \, Selection})$ at a fixed mass is $\sim0.1$ dex to $\sim0.6$ dex. The tension increases with mass and redshift.  The large tension with the normalizations and exponents of SFR$-{\rm M}_{\star}$ relations from Lyman-break, SFR and H$\alpha$ selected samples at all redshifts suggests that Lyman-break and H$\alpha$ selections could be biased and possibly do not take into account a significant number of objects. Cosmological simulations predict steeper slopes than observations, with an exponent close to unity almost at all redshifts. In general, numerical results from different groups are in good agreement with mass selected observations. Recent relations that take into account fainter objects are significantly steeper than those found by past authors and this is in agreement with the results of \citet{reddy2012}, \citet{Guo2013} and the predictions from cosmological simulations \citep{Katsianis2014}. Simulations implemented with variable energy driven and momentum driven winds give similar results for the   SFR-$-{\rm M}_{\star}$ relation and are able to reproduce the observables, while models with constant winds fail to produce realistic results at low redshift.

\section{Discussion and conclusions}
\label{Disc}

In the previous sections we have investigated the evolution of the SFR$-{\rm M}_{\star}$ relation for galaxies at $z \sim1-4$ using cosmological hydrodynamic simulations and a compilation of observed relations from various groups. In particular, we demonstrated that observational studies report a range of SFR$-{\rm M}_{\star}$ relations at high redshifts. The reason for the lack of consensus is probably  related to the fact that different groups adopt different methods of selection and calculation of the intrinsic properties of galaxies. There has been a considerable effort to constrain the observed star formation rates and stellar masses of galaxies but there is not yet a conclusive method of  measurements. This has to be addressed by future observers since different methods produce different results. In addition, in the previous sections we demonstrated that the SFR$-{\rm M}_{\star}$ relations obtained from Lyman-break selected and star forming galaxies have typically  higher normalizations, when compared with stellar mass based selection  and cosmological hydrodynamic simulations from various groups. This can be  due to the fact that these samples are probably biased to include the most luminous-star forming objects and possibly miss a large population of low SFR galaxies. This tends to increase the observed mean SFR at a fixed mass. The tension between different groups becomes smaller at lower redshifts, where observations are more complete.  This is possibly due to the fact that various biases related to sample selection and limits of instrumentation become less severe. Overall, we find that simulations are in good agreement with studies which use SED fitting to estimate the intrinsic SFRs, dust corrections and stellar masses for the observed objects. However, models are unable to reproduce SFR(UV+IR)$-{\rm M}_{\star}$  relations.

A comparison between different simulations at $z \sim 1-4$ suggests that the assumed parameters for the AGN and variable winds feedback implementations  do not affect the simulated SFR$-{\rm M}_{\star}$ relation within the range of models considered. While more extreme models can produce different SFR$-{\rm M}_{\star}$ relations \citep{Haas2013}, the models in this paper were chosen to reproduce the observed SFRF and GSMF at redshifts $z \sim 1-7$ \citep{TescariKaW2013,Katsianis2014}. We find that the scatter of the SFR$-{\rm M}_{\star}$ relation is $ \sim0.2$ dex at all redshifts, which is in agreement with estimates from recent observations \citep{Whitaker2012,Speagle14}.

We note that the implementation of star formation processes and ISM physics in simulations has remained essentially unchanged in the last decade. This could be the reason of the surprising agreement between older simulations and new simulations with an order of magnitude higher resolutions and larger box sizes. Future numerical codes should aim to improve the ISM modeling in order to provide more robust estimations of, among other galaxy properties, the redshift evolution of the SFR$-{\rm M}_{\star}$ relation.

\section*{Acknowledgments} 

The authors would like to thank Amanda Bauer, Naveen Reddy, Masaru Kajisawa and Yusei Koyama for insightful discussions and for kindly sharing their observational  results with us. In addition, we would like to thank Martin Sparre and Michelle Furlong for sharing their numerical results. We would also like to thank Volker Springel for making available to us the non-public version of the {\small{GADGET-3}} code.    Finally, we are particularly grateful to Mark Sargent for his suggestions. This research was conducted by the Australian Research Council Centre of Excellence for All-sky Astrophysics (CAASTRO), through project number CE110001020. This work was supported by the NCI National Facility at the ANU, the Melbourne International Research Scholarship (MIRS) scholarship and the Albert Shimmins Fund - writing-up award provided by the University of Melbourne. AK is supported by the CONICYT-FONDECYT fellowship (project number: 3160049). 


\bibliographystyle{apj} 
\bibliography{pasa-sample} 

\newcommand{\noopsort}[1]{}
\begin{thebibliography}{65}
\expandafter\ifx\csname natexlab\endcsname\relax\def\natexlab#1{#1}\fi

\bibitem[{{Barai} {et~al.}(2013){Barai}, {Viel}, {Borgani}, {Tescari},
  {Tornatore}, {Dolag}, {Killedar}, {Monaco}, {D'Odorico}, \&
  {Cristiani}}]{barai13}
{Barai}, P., {Viel}, M., {Borgani}, S., {Tescari}, E., {Tornatore}, L.,
  {Dolag}, K., {Killedar}, M., {Monaco}, P., {D'Odorico}, V., \& {Cristiani},
  S. 2013, \mnras, 430, 3213

\bibitem[{{Bauer} {et~al.}(2011){Bauer}, {Conselice}, {P{\'e}rez-Gonz{\'a}lez},
  {Gr{\"u}tzbauch}, {Bluck}, {Buitrago}, \& {Mortlock}}]{Bauer11}
{Bauer}, A.~E., {Conselice}, C.~J., {P{\'e}rez-Gonz{\'a}lez}, P.~G.,
  {Gr{\"u}tzbauch}, R., {Bluck}, A.~F.~L., {Buitrago}, F., \& {Mortlock}, A.
  2011, \mnras, 417, 289

\bibitem[{{Bauer} {et~al.}(2013){Bauer}, {Hopkins}, {Gunawardhana},
  {et~al.}}]{Bauer2013}
{Bauer}, A.~E., {Hopkins}, A.~M., {Gunawardhana}, M., {et~al.} 2013, \mnras,
  434, 209

\bibitem[{{Behroozi} {et~al.}(2013){Behroozi}, {Wechsler}, \&
  {Conroy}}]{Behroozi2013}
{Behroozi}, P.~S., {Wechsler}, R.~H., \& {Conroy}, C. 2013, \apj, 770, 57

\bibitem[{{Boquien} {et~al.}(2014){Boquien}, {Buat}, \& {Perret}}]{Boquien2014}
{Boquien}, M., {Buat}, V., \& {Perret}, V. 2014, \aap, 571, A72

\bibitem[{{Bouwens} {et~al.}(2012){Bouwens}, {Illingworth}, {Oesch}, {Franx},
  {Labb{\'e}}, {Trenti}, {van Dokkum}, {Carollo}, {Gonz{\'a}lez}, {Smit}, \&
  {Magee}}]{bouwens2012}
{Bouwens}, R.~J., {Illingworth}, G.~D., {Oesch}, P.~A., {Franx}, M.,
  {Labb{\'e}}, I., {Trenti}, M., {van Dokkum}, P., {Carollo}, C.~M.,
  {Gonz{\'a}lez}, V., {Smit}, R., \& {Magee}, D. 2012, \apj, 754, 83

\bibitem[{{Brinchmann} \& {Ellis}(2000)}]{Brinchmann00}
{Brinchmann}, J. \& {Ellis}, R.~S. 2000, \apjl, 536, L77

\bibitem[{{Broadhurst} {et~al.}(1992){Broadhurst}, {Ellis}, \&
  {Glazebrook}}]{Broad92}
{Broadhurst}, T.~J., {Ellis}, R.~S., \& {Glazebrook}, K. 1992, \nat, 355, 55

\bibitem[{{Bruzual} \& {Charlot}(2003)}]{Bruzualch03}
{Bruzual}, G. \& {Charlot}, S. 2003, \mnras, 344, 1000

\bibitem[{{Calzetti}(1997)}]{Calzetti1997}
{Calzetti}, D. 1997, \aj, 113, 162

\bibitem[{{Calzetti} {et~al.}(2000){Calzetti}, {Armus}, {Bohlin}, {Kinney},
  {Koornneef}, \& {Storchi-Bergmann}}]{Calzetti2000}
{Calzetti}, D., {Armus}, L., {Bohlin}, R.~C., {Kinney}, A.~L., {Koornneef}, J.,
  \& {Storchi-Bergmann}, T. 2000, \apj, 533, 682

\bibitem[{{Chabrier}(2003)}]{Chabrier03}
{Chabrier}, G. 2003, \pasp, 115, 763

\bibitem[{{Crain} {et~al.}(2015){Crain}, {Schaye}, {Bower}, {Furlong},
  {Schaller}, {Theuns}, {Dalla Vecchia}, {Frenk}, {McCarthy}, {Helly},
  {Jenkins}, {Rosas-Guevara}, {White}, \& {Trayford}}]{Crain2015}
{Crain}, R.~A., {Schaye}, J., {Bower}, R.~G., {Furlong}, M., {Schaller}, M.,
  {Theuns}, T., {Dalla Vecchia}, C., {Frenk}, C.~S., {McCarthy}, I.~G.,
  {Helly}, J.~C., {Jenkins}, A., {Rosas-Guevara}, Y.~M., {White}, S.~D.~M., \&
  {Trayford}, J.~W. 2015, \mnras, 450, 1937

\bibitem[{{Daddi} {et~al.}(2007){Daddi}, {Dickinson}, {Morrison}, {Chary},
  {Cimatti}, {Elbaz}, {Frayer}, {Renzini}, {Pope}, {Alexander}, {Bauer},
  {Giavalisco}, {Huynh}, {Kurk}, \& {Mignoli}}]{Daddi2007}
{Daddi}, E., {Dickinson}, M., {Morrison}, G., {Chary}, R., {Cimatti}, A.,
  {Elbaz}, D., {Frayer}, D., {Renzini}, A., {Pope}, A., {Alexander}, D.~M.,
  {Bauer}, F.~E., {Giavalisco}, M., {Huynh}, M., {Kurk}, J., \& {Mignoli}, M.
  2007, \apj, 670, 156

\bibitem[{{Dav{\'e}}(2008)}]{Dave08}
{Dav{\'e}}, R. 2008, \mnras, 385, 147

\bibitem[{{Dayal} \& {Ferrara}(2012)}]{DayalFerrara2012}
{Dayal}, P. \& {Ferrara}, A. 2012, \mnras, 421, 2568

\bibitem[{{de Barros} {et~al.}(2014){de Barros}, {Schaerer}, \&
  {Stark}}]{deBarros}
{de Barros}, S., {Schaerer}, D., \& {Stark}, D.~P. 2014, \aap, 563, A81

\bibitem[{{de los Reyes} {et~al.}(2015){de los Reyes}, {Ly}, {Lee}, {Salim},
  {Peeples}, {Momcheva}, {Feddersen}, {Dale}, {Ouchi}, {Ono}, \&
  {Finn}}]{Delosrayes2014}
{de los Reyes}, M.~A., {Ly}, C., {Lee}, J.~C., {Salim}, S., {Peeples}, M.~S.,
  {Momcheva}, I., {Feddersen}, J., {Dale}, D.~A., {Ouchi}, M., {Ono}, Y., \&
  {Finn}, R. 2015, \aj, 149, 79

\bibitem[{{Drory} \& {Alvarez}(2008)}]{Drory08}
{Drory}, N. \& {Alvarez}, M. 2008, \apj, 680, 41

\bibitem[{{Drory} {et~al.}(2005){Drory}, {Salvato}, {Gabasch}, {Bender},
  {Hopp}, {Feulner}, \& {Pannella}}]{Drory2005}
{Drory}, N., {Salvato}, M., {Gabasch}, A., {Bender}, R., {Hopp}, U., {Feulner},
  G., \& {Pannella}, M. 2005, \apjl, 619, L131

\bibitem[{{Dutton} {et~al.}(2010){Dutton}, {van den Bosch}, \&
  {Dekel}}]{Dutton10}
{Dutton}, A.~A., {van den Bosch}, F.~C., \& {Dekel}, A. 2010, \mnras, 405, 1690

\bibitem[{{Elbaz} {et~al.}(2007){Elbaz}, {Daddi}, {Le Borgne}, {Dickinson},
  {Alexander}, {Chary}, {Starck}, {Brandt}, {Kitzbichler}, {MacDonald},
  {Nonino}, {Popesso}, {Stern}, \& {Vanzella}}]{Elbaz07}
{Elbaz}, D., {Daddi}, E., {Le Borgne}, D., {Dickinson}, M., {Alexander}, D.~M.,
  {Chary}, R.-R., {Starck}, J.-L., {Brandt}, W.~N., {Kitzbichler}, M.,
  {MacDonald}, E., {Nonino}, M., {Popesso}, P., {Stern}, D., \& {Vanzella}, E.
  2007, \aap, 468, 33

\bibitem[{{Ferland} {et~al.}(2013){Ferland}, {Porter}, {van Hoof}, {Williams},
  {Abel}, {Lykins}, {Shaw}, {Henney}, \& {Stancil}}]{ferland13}
{Ferland}, G.~J., {Porter}, R.~L., {van Hoof}, P.~A.~M., {Williams}, R.~J.~R.,
  {Abel}, N.~P., {Lykins}, M.~L., {Shaw}, G., {Henney}, W.~J., \& {Stancil},
  P.~C. 2013, arxiv:1302.448

\bibitem[{{Finlator} {et~al.}(2011){Finlator}, {Dav{\'e}}, \&
  {{\"O}zel}}]{Finlator11}
{Finlator}, K., {Dav{\'e}}, R., \& {{\"O}zel}, F. 2011, \apj, 743, 169

\bibitem[{{Fumagalli} {et~al.}(2014){Fumagalli}, {Labb{\'e}}, {Patel}, {Franx},
  {van Dokkum}, {Brammer}, {da Cunha}, {F{\"o}rster Schreiber}, {Kriek},
  {Quadri}, {Rix}, {Wake}, {Whitaker}, {Lundgren}, {Marchesini}, {Maseda},
  {Momcheva}, {Nelson}, {Pacifici}, \& {Skelton}}]{Fumagalli2014}
{Fumagalli}, M., {Labb{\'e}}, I., {Patel}, S.~G., {Franx}, M., {van Dokkum},
  P., {Brammer}, G., {da Cunha}, E., {F{\"o}rster Schreiber}, N.~M., {Kriek},
  M., {Quadri}, R., {Rix}, H.-W., {Wake}, D., {Whitaker}, K.~E., {Lundgren},
  B., {Marchesini}, D., {Maseda}, M., {Momcheva}, I., {Nelson}, E., {Pacifici},
  C., \& {Skelton}, R.~E. 2014, \apj, 796, 35

\bibitem[{{Furlong} {et~al.}(2015){Furlong}, {Bower}, {Theuns}, {Schaye},
  {Crain}, {Schaller}, {Dalla Vecchia}, {Frenk}, {McCarthy}, {Helly},
  {Jenkins}, \& {Rosas-Guevara}}]{Furlong2014}
{Furlong}, M., {Bower}, R.~G., {Theuns}, T., {Schaye}, J., {Crain}, R.~A.,
  {Schaller}, M., {Dalla Vecchia}, C., {Frenk}, C.~S., {McCarthy}, I.~G.,
  {Helly}, J., {Jenkins}, A., \& {Rosas-Guevara}, Y.~M. 2015, \mnras, 450, 4486

\bibitem[{{Garn} \& {Best}(2010)}]{Garn2010}
{Garn}, T. \& {Best}, P.~N. 2010, \mnras, 409, 421

\bibitem[{{Genel} {et~al.}(2014){Genel}, {Vogelsberger}, {Springel}, {Sijacki},
  {Nelson}, {Snyder}, {Rodriguez-Gomez}, {Torrey}, \& {Hernquist}}]{Genel2014}
{Genel}, S., {Vogelsberger}, M., {Springel}, V., {Sijacki}, D., {Nelson}, D.,
  {Snyder}, G., {Rodriguez-Gomez}, V., {Torrey}, P., \& {Hernquist}, L. 2014,
  \mnras, 445, 175

\bibitem[{{Gilbank} {et~al.}(2011){Gilbank}, {Bower}, {Glazebrook}, {Balogh},
  {Baldry}, {Davies}, {Hau}, {Li}, {McCarthy}, \& {Sawicki}}]{Gilbank2011}
{Gilbank}, D.~G., {Bower}, R.~G., {Glazebrook}, K., {Balogh}, M.~L., {Baldry},
  I.~K., {Davies}, G.~T., {Hau}, G.~K.~T., {Li}, I.~H., {McCarthy}, P., \&
  {Sawicki}, M. 2011, \mnras, 414, 304

\bibitem[{{Guo} {et~al.}(2013){Guo}, {Zheng}, \& {Fu}}]{Guo2013}
{Guo}, K., {Zheng}, X.~Z., \& {Fu}, H. 2013, \apj, 778, 23

\bibitem[{{Haardt} \& {Madau}(2001)}]{haardtmadau01}
{Haardt}, F. \& {Madau}, P. 2001, in Clusters of Galaxies and the High Redshift
  Universe Observed in X-rays, ed. D.~M. {Neumann} \& J.~T.~V. {Tran}

\bibitem[{{Haas} {et~al.}(2013){Haas}, {Schaye}, {Booth}, {Dalla Vecchia},
  {Springel}, {Theuns}, \& {Wiersma}}]{Haas2013}
{Haas}, M.~R., {Schaye}, J., {Booth}, C.~M., {Dalla Vecchia}, C., {Springel},
  V., {Theuns}, T., \& {Wiersma}, R.~P.~C. 2013, \mnras, 435, 2955

\bibitem[{{Hayward} {et~al.}(2014){Hayward}, {Lanz}, {Ashby}, {Fazio},
  {Hernquist}, {Mart{\'{\i}}nez-Galarza}, {Noeske}, {Smith}, {Wuyts}, \&
  {Zezas}}]{Hayward2014}
{Hayward}, C.~C., {Lanz}, L., {Ashby}, M.~L.~N., {Fazio}, G., {Hernquist}, L.,
  {Mart{\'{\i}}nez-Galarza}, J.~R., {Noeske}, K., {Smith}, H.~A., {Wuyts}, S.,
  \& {Zezas}, A. 2014, \mnras, 445, 1598

\bibitem[{{Heinis} {et~al.}(2014){Heinis}, {Buat}, {B{\'e}thermin}, {Bock},
  {et~al.}}]{Heinis2014}
{Heinis}, S., {Buat}, V., {B{\'e}thermin}, M., {Bock}, J., {et~al.} 2014,
  \mnras, 437, 1268

\bibitem[{{Kajisawa} {et~al.}(2010){Kajisawa}, {Ichikawa}, {Yamada},
  {Uchimoto}, {Yoshikawa}, {Akiyama}, \& {Onodera}}]{Kajisawa2010}
{Kajisawa}, M., {Ichikawa}, T., {Yamada}, T., {Uchimoto}, Y.~K., {Yoshikawa},
  T., {Akiyama}, M., \& {Onodera}, M. 2010, \apj, 723, 129

\bibitem[{{Kannan} {et~al.}(2014){Kannan}, {Stinson}, {Macci{\`o}}, {Brook},
  {Weinmann}, {Wadsley}, \& {Couchman}}]{Kannan2014}
{Kannan}, R., {Stinson}, G.~S., {Macci{\`o}}, A.~V., {Brook}, C., {Weinmann},
  S.~M., {Wadsley}, J., \& {Couchman}, H.~M.~P. 2014, \mnras, 437, 3529

\bibitem[{{Karim} {et~al.}(2011){Karim}, {Schinnerer},
  {Mart{\'{\i}}nez-Sansigre}, \& others.}]{Karim2011}
{Karim}, A., {Schinnerer}, E., {Mart{\'{\i}}nez-Sansigre}, A., \& others. 2011,
  \apj, 730, 61

\bibitem[{{Kashino} {et~al.}(2013){Kashino}, {Silverman}, {Rodighiero},
  {et~al.}}]{Kashino2013}
{Kashino}, D., {Silverman}, J.~D., {Rodighiero}, G., {et~al.} 2013, \apjl, 777,
  L8

\bibitem[{{Katsianis} {et~al.}(2015){Katsianis}, {Tescari}, \&
  {Wyithe}}]{Katsianis2014}
{Katsianis}, A., {Tescari}, E., \& {Wyithe}, J.~S.~B. 2015, \mnras, 448, 3001

\bibitem[{{Kennicutt}(1998)}]{kennicutt1998}
{Kennicutt}, Jr., R.~C. 1998, \araa, 36, 189

\bibitem[{{Koyama} {et~al.}(2013){Koyama}, {Smail}, {Kurk}, {Geach}, {Sobral},
  {Kodama}, {Nakata}, {Swinbank}, {Best}, {Hayashi}, \& {Tadaki}}]{Koyama2013}
{Koyama}, Y., {Smail}, I., {Kurk}, J., {Geach}, J.~E., {Sobral}, D., {Kodama},
  T., {Nakata}, F., {Swinbank}, A.~M., {Best}, P.~N., {Hayashi}, M., \&
  {Tadaki}, K.-i. 2013, \mnras, 434, 423

\bibitem[{{Kroupa}(2001)}]{kroupa01}
{Kroupa}, P. 2001, \mnras, 322, 231

\bibitem[{{Magdis} {et~al.}(2010){Magdis}, {Rigopoulou}, {Huang}, \&
  {Fazio}}]{Magdis10}
{Magdis}, G.~E., {Rigopoulou}, D., {Huang}, J.-S., \& {Fazio}, G.~G. 2010,
  \mnras, 401, 1521

\bibitem[{{Meurer} {et~al.}(1999){Meurer}, {Heckman}, \&
  {Calzetti}}]{meurer1999}
{Meurer}, G.~R., {Heckman}, T.~M., \& {Calzetti}, D. 1999, \apj, 521, 64

\bibitem[{{Murante} {et~al.}(2015){Murante}, {Monaco}, {Borgani}, {Tornatore},
  {Dolag}, \& {Goz}}]{Murante2015}
{Murante}, G., {Monaco}, P., {Borgani}, S., {Tornatore}, L., {Dolag}, K., \&
  {Goz}, D. 2015, \mnras, 447, 178

\bibitem[{{Noeske} {et~al.}(2007)}]{Noeske2007}
{Noeske}, K.~G. {et~al.} 2007, \apjl, 660, L43

\bibitem[{{Pei}(1992)}]{Pei1992}
{Pei}, Y.~C. 1992, \apj, 395, 130

\bibitem[{{Puchwein} \& {Springel}(2013)}]{PuchweinSpri12}
{Puchwein}, E. \& {Springel}, V. 2013, \mnras, 428, 2966

\bibitem[{{Reddy} {et~al.}(2012){Reddy}, {Pettini}, {Steidel}, {Shapley},
  {Erb}, \& {Law}}]{reddy2012}
{Reddy}, N.~A., {Pettini}, M., {Steidel}, C.~C., {Shapley}, A.~E., {Erb},
  D.~K., \& {Law}, D.~R. 2012, \apj, 754, 25

\bibitem[{{Salmon} {et~al.}(2015){Salmon}, {Papovich}, {Finkelstein}, {Tilvi},
  {Finlator}, {Behroozi}, {Dahlen}, {Dav{\'e}}, {Dekel}, {Dickinson},
  {Ferguson}, {Giavalisco}, {Long}, {Lu}, {Mobasher}, {Reddy}, {Somerville}, \&
  {Wechsler}}]{Salmon2015}
{Salmon}, B., {Papovich}, C., {Finkelstein}, S.~L., {Tilvi}, V., {Finlator},
  K., {Behroozi}, P., {Dahlen}, T., {Dav{\'e}}, R., {Dekel}, A., {Dickinson},
  M., {Ferguson}, H.~C., {Giavalisco}, M., {Long}, J., {Lu}, Y., {Mobasher},
  B., {Reddy}, N., {Somerville}, R.~S., \& {Wechsler}, R.~H. 2015, \apj, 799,
  183

\bibitem[{{Salpeter}(1955)}]{salpeter55}
{Salpeter}, E.~E. 1955, \apj, 121, 161

\bibitem[{{Schaye} {et~al.}(2015){Schaye}, {Crain}, {Bower}, {Furlong},
  {et~al.}}]{Schaye2015}
{Schaye}, J., {Crain}, R.~A., {Bower}, R.~G., {Furlong}, M., {et~al.} 2015,
  \mnras, 446, 521

\bibitem[{{Sobral} {et~al.}(2013){Sobral}, {Smail}, {Best}, {Geach}, {Matsuda},
  {Stott}, {Cirasuolo}, \& {Kurk}}]{Sobral2013}
{Sobral}, D., {Smail}, I., {Best}, P.~N., {Geach}, J.~E., {Matsuda}, Y.,
  {Stott}, J.~P., {Cirasuolo}, M., \& {Kurk}, J. 2013, \mnras, 428, 1128

\bibitem[{{Sparre} {et~al.}(2015){Sparre}, {Hayward}, {Springel},
  {Vogelsberger}, {Genel}, {Torrey}, {Nelson}, {Sijacki}, \&
  {Hernquist}}]{Sparre2014}
{Sparre}, M., {Hayward}, C.~C., {Springel}, V., {Vogelsberger}, M., {Genel},
  S., {Torrey}, P., {Nelson}, D., {Sijacki}, D., \& {Hernquist}, L. 2015,
  \mnras, 447, 3548

\bibitem[{{Speagle} {et~al.}(2014){Speagle}, {Steinhardt}, {Capak}, \&
  {Silverman}}]{Speagle14}
{Speagle}, J.~S., {Steinhardt}, C.~L., {Capak}, P.~L., \& {Silverman}, J.~D.
  2014, arXiv:1405.2041

\bibitem[{{Springel}(2005)}]{springel2005}
{Springel}, V. 2005, \mnras, 364, 1105

\bibitem[{{Springel} \& {Hernquist}(2003)}]{springel2003}
{Springel}, V. \& {Hernquist}, L. 2003, \mnras, 339, 289

\bibitem[{{Tescari} {et~al.}(2014){Tescari}, {Katsianis}, {Wyithe}, {Dolag},
  {Tornatore}, {Barai}, {Viel}, \& {Borgani}}]{TescariKaW2013}
{Tescari}, E., {Katsianis}, A., {Wyithe}, J.~S.~B., {Dolag}, K., {Tornatore},
  L., {Barai}, P., {Viel}, M., \& {Borgani}, S. 2014, \mnras, 438, 3490

\bibitem[{{Tomczak} {et~al.}(2016){Tomczak}, {Quadri}, {Tran}, {Labb{\'e}},
  {Straatman}, {Papovich}, {Glazebrook}, {Allen}, {Brammer}, {Cowley},
  {Dickinson}, {Elbaz}, {Inami}, {Kacprzak}, {Morrison}, {Nanayakkara},
  {Persson}, {Rees}, {Salmon}, {Schreiber}, {Spitler}, \&
  {Whitaker}}]{Tomczak2016}
{Tomczak}, A.~R., {Quadri}, R.~F., {Tran}, K.-V.~H., {Labb{\'e}}, I.,
  {Straatman}, C.~M.~S., {Papovich}, C., {Glazebrook}, K., {Allen}, R.,
  {Brammer}, G.~B., {Cowley}, M., {Dickinson}, M., {Elbaz}, D., {Inami}, H.,
  {Kacprzak}, G.~G., {Morrison}, G.~E., {Nanayakkara}, T., {Persson}, S.~E.,
  {Rees}, G.~A., {Salmon}, B., {Schreiber}, C., {Spitler}, L.~R., \&
  {Whitaker}, K.~E. 2016, \apj, 817, 118

\bibitem[{{Tornatore} {et~al.}(2007){Tornatore}, {Borgani}, {Dolag}, \&
  {Matteucci}}]{T07}
{Tornatore}, L., {Borgani}, S., {Dolag}, K., \& {Matteucci}, F. 2007, \mnras,
  382, 1050

\bibitem[{{Utomo} {et~al.}(2014){Utomo}, {Kriek}, {Labb{\'e}}, {Conroy}, \&
  {Fumagalli}}]{Utomo2014}
{Utomo}, D., {Kriek}, M., {Labb{\'e}}, I., {Conroy}, C., \& {Fumagalli}, M.
  2014, \apjl, 783, L30

\bibitem[{{Vogelsberger} {et~al.}(2013){Vogelsberger}, {Genel}, {Sijacki},
  {Torrey}, {Springel}, \& {Hernquist}}]{Vogelsberger2013}
{Vogelsberger}, M., {Genel}, S., {Sijacki}, D., {Torrey}, P., {Springel}, V.,
  \& {Hernquist}, L. 2013, \mnras, 436, 3031

\bibitem[{{Whitaker} {et~al.}(2014){Whitaker}, {Franx}, {Leja}, {van Dokkum},
  {Henry}, {Skelton}, {Fumagalli}, {Momcheva}, {Brammer}, {Labb{\'e}},
  {Nelson}, \& {Rigby}}]{Whitaker2014}
{Whitaker}, K.~E., {Franx}, M., {Leja}, J., {van Dokkum}, P.~G., {Henry}, A.,
  {Skelton}, R.~E., {Fumagalli}, M., {Momcheva}, I.~G., {Brammer}, G.~B.,
  {Labb{\'e}}, I., {Nelson}, E.~J., \& {Rigby}, J.~R. 2014, \apj, 795, 104

\bibitem[{{Whitaker} {et~al.}(2012){Whitaker}, {van Dokkum}, {Brammer}, \&
  {Franx}}]{Whitaker2012}
{Whitaker}, K.~E., {van Dokkum}, P.~G., {Brammer}, G., \& {Franx}, M. 2012,
  \apjl, 754, L29

\bibitem[{{Wiersma} {et~al.}(2009){Wiersma}, {Schaye}, \& {Smith}}]{wiersma09}
{Wiersma}, R.~P.~C., {Schaye}, J., \& {Smith}, B.~D. 2009, \mnras, 393, 99

\end{thebibliography}

\begin{appendix}

\section{Have past observed SFR$-{\rm M}_{\star}$ relations misguided simulations?}
\label{Comparisonold}

\begin{figure*}[!t]
\centering
\vspace{0.0cm}
\hspace{0.0cm}
\includegraphics[scale=0.73]{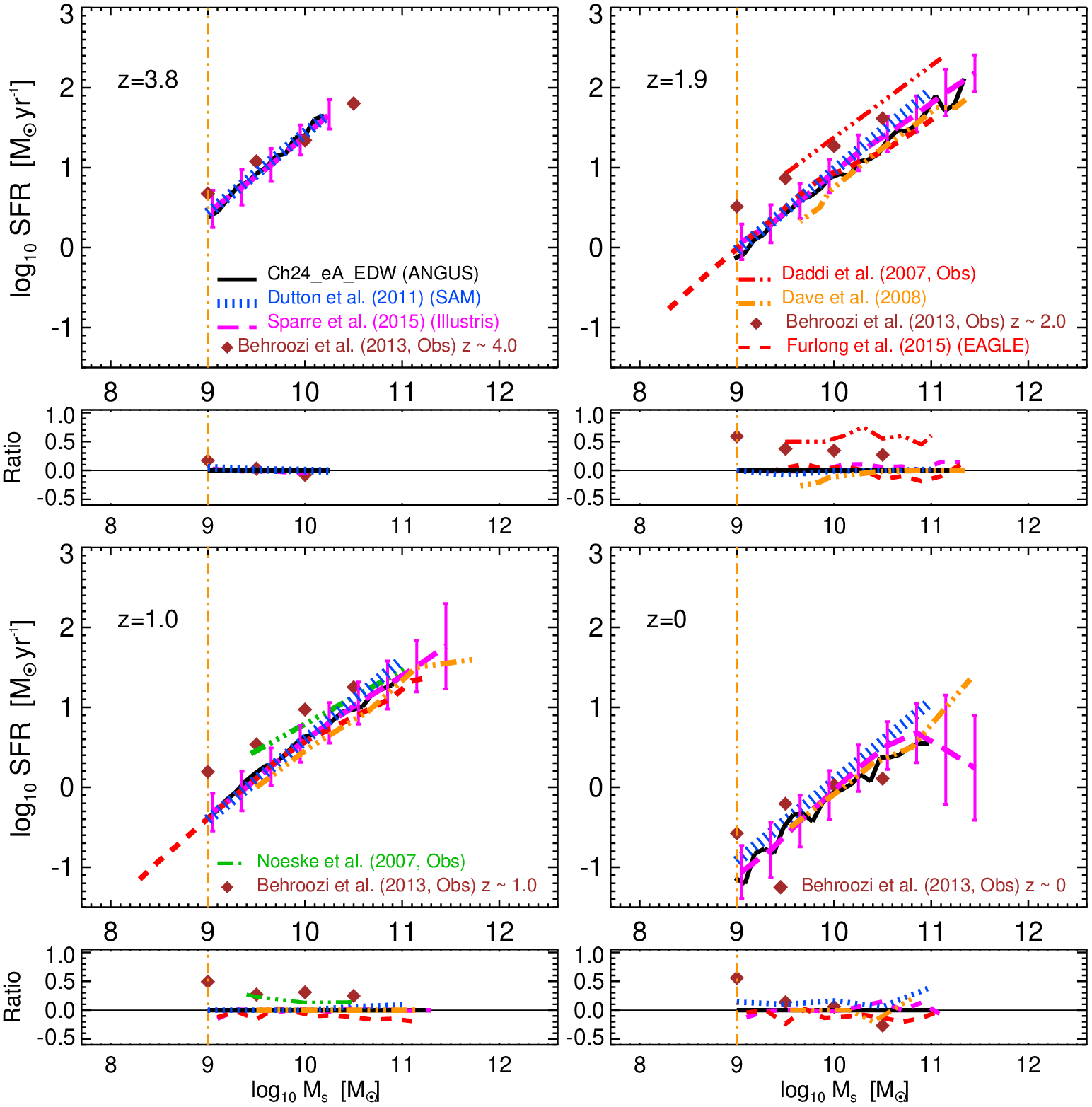}
\vspace{0.32cm}
\caption{Median values of the SFR$-{\rm M}_{\star}$ relations from different cosmological hydrodynamic simulations for $z \sim 0 - 4$. The black line is our reference model (\textit{Ch24\tu eA\tu EDW}).  The dark green dotted line is the median fit of the scatter plot presented by \citet{Dave08}. The blue dotted line is the median fit of the scatter plot presented by \citet[][SAM]{Dutton10}. The magenta dashed line is the median line of the scatter plot presented by \citet[][Illustris project]{Sparre2014}. The red dashed line is the median line of the scatter plot presented by \citet[][EAGLE project]{Furlong2014}. We cut our ${\rm SFR}({\rm M}_{\star})$  under our confidence mass limit of $10^{9}$ M$_{\rm \odot}$ to make a meaningful comparison with the Illustris and EAGLE projects. There is an excellent agreement between the results from cosmological hydrodynamic simulations run by different groups. At each redshift, a panel showing ratios between the different simulations and observations with the \textit{Ch24\tu eA\tu EDW} (black solid line) is included.}
\label{fig:CompSFR2}
\end{figure*}

In this appendix we discuss how the uncertainty of dust correction laws and selection effects/biases have been affecting the comparison with cosmological simulations for the past decade. In figure \ref{fig:CompSFR2} we show a compilation of median SFR$-{\rm M}_{\star}$ relations from cosmological hydrodynamic simulations and semi-analytic models alongside with observations from various groups.

\citet{Dave08} investigated the tension between observed and simulated results for the SFR$-{\rm M}_{\star}$ and extensively discussed how to address it. In general, the numerical results implied steeper relations with lower normalization than observations. The author considered various modifications of the theoretical picture of stellar mass assembly, but each was found to be in conflict with observations of high-redshift galaxies. In light of this tension, \citet{Dave08} suggested an evolving IMF to address the discrepancy. Is it possible though, that the observations used \citep[e.g. ][]{Noeske2007,Daddi2007} to constrain the simulations contained biases or/and overestimated the SFR at fixed stellar mass? For redshift $z \sim 0$, we see that the numerical results of \citet{Dave08} are in excellent agreement with current state-of-the-art cosmological simulations from the Illustris and EAGLE projects, despite of the fact that the resolution is higher and box sizes are larger for the latter two simulations. At redshift $z \sim1$, the slope of the simulated relation presented by \citet{Dave08} is steeper than the observations of \citet{Noeske2007}. However, in the previous sections we saw indications that the results of \citet{Noeske2007} could have been artificially shallower due to the fact that they were not taking into acount low mass/SFR objects. The updated results from \citet{Guo2013} that take into account low SFR objects and $24 \mu m$ undetected galaxies, are significantly steeper with a power law exponent close to unity. Once again, at $z \sim1$ we see the perfect agreement between the numerical results of the Illustris and EAGLE projects and the simulated relation from \citet{Dave08}.  Moving to $z \sim2$, \citet{Dave08} reported a significant tension, with an amplitude offset of $\sim4-5$, with the results of \citet{Daddi2007}. In the above sections we saw that the relation given by \citet{Daddi2007} could have an  artificially high normalization and shallow slope \citep{Bauer11,Hayward2014}, since the authors preferably selected star forming galaxies and relied on SFRs that were obtained from UV+IR luminosities. The numerical results of \citet{Dave08} are in good agreement with the simulated SFR$-{\rm M}_{\star}$ from the  {\textsc{Angus}}, Illustris and EAGLE projects, and the mass selected observations that adopt SED fitting techniques to obtain dust corrections and SFRs. Maybe the tension between observed and simulated relations reported by \citet{Dave08} could have its roots in the fact that past observations \citep{Noeske2007,Daddi2007} were overestimating the SFR at a fixed stellar mass due to methodology, sample selection effects and biases related to undetected faint objects.

\citet{Sparre2014} stated that there is good consistency between the simulated and observed \citep{Behroozi2013} SFR$-{\rm M}_{\star}$ relations at $z \sim0$ and $z \sim4$. However, at intermediate redshifts there is a severe tension. We stress  that while the results of \citet{Behroozi2013} are a compilation of observations, the authors did not account for the fact that these observations assumed completely different methods to produce SFR$-{\rm M}_{\star}$ relations. Different methods produce completely different results  and for this reason they should not be compiled directly all together. The results of \citet{Behroozi2013} for redshifts $z \sim1-2$ were mostly based on samples that preferably selected star forming galaxies and assumed IR+UV luminosities to obtain the intrinsic SFRs  \citep{Noeske2007,Daddi2007,Whitaker2012}. Therefore, the compilation could overpredict the SFR at a fixed stellar mass for these redshifts, with respect to mass selected surveys that used detailed SED fitting. In Figure \ref{fig:CompSFR2} we show that all models underpredict the SFR at a fixed stellar mass for $z \sim1-2$, with respect to the compilation of observations in \citet{Behroozi2013}. The comparison suggests that it is quite possible that the compilation did not take into account the faint objects at $z \sim1-2$ and therefore suffers from the Malmquist bias (the difference with the intrinsic relation suggested by simulations increases at lower masses). A correction to lower star formation rates would be expected, as in \citet{reddy2012}, in order to obtain an unbiased relation due to incompleteness. This would bring observations and simulations to better agreement.

In conclusion, the tension reported  in the literature likely has its roots in the fact that the comparisons have been done using relations that possibly overpredict the SFR at a fixed stellar mass due to selection biases and/or methodology. Cosmological simulations in the past decade indicate that the slope of the relation is steeper with a lower normalization. Of course, it is possible that this slope is the result of a poor representation of physical processes, which are implemented in a similar way by simulators ({\textsc{Angus}}, EAGLE, Illustris). We demonstrated that the simulated SFR$-{\rm M}_{\star}$ relations from various groups are in excellent agreement and largely independent of resolution and box size. This somewhat surprising result points out how more work is needed to improve the numerical modeling of star formation processes and, in particular, the description of the ISM.

\end{appendix}

\end{document}